\documentclass[11pt, a4paper]{article}

\usepackage[utf8]{inputenc}
\usepackage[english]{babel}
\usepackage{times} % Verwendet eine gängige Schriftart wie Times New Roman
\usepackage{csquotes}

\usepackage{amsmath}
\usepackage{amsfonts}
\usepackage{amssymb}

\usepackage{graphicx} % Für das Einbinden von Bildern

\usepackage{booktabs} % Für professionelle Tabellen (\toprule, \midrule, \bottomrule)
\usepackage{longtable} % Für Tabellen, die über eine Seite hinausgehen
\usepackage{tabularx} 
\usepackage{ltablex} 
\keepXColumns
\usepackage{ragged2e} 
\usepackage{array} 
\usepackage{mwe} %

\usepackage{xcolor}   
\usepackage{listings}  

\usepackage{titling}
\usepackage{titlesec}
\usepackage[toc,page]{appendix}

\pretitle{\begin{center}\LARGE}
\posttitle{\par\end{center}\vspace{-0.6em}}
\preauthor{\begin{center}\small}
\postauthor{\par\end{center}\vspace{-0.4em}}
\predate{\begin{center}\small}
\postdate{\par\end{center}\vspace{-0.8em}}

\lstdefinestyle{mystyle}{
    backgroundcolor=\color{black!5},
    commentstyle=\color{green!40!black},
    keywordstyle=\color{blue},
    numberstyle=\tiny\color{black!50},
    stringstyle=\color{purple},
    basicstyle=\footnotesize\ttfamily,
    breakatwhitespace=false,         
    breaklines=true,                 
    captionpos=b,                    
    keepspaces=true,                 
    numbers=left,                    
    numbersep=5pt,                  
    showspaces=false,                
    showstringspaces=false,
    showtabs=false,                  
    tabsize=2
}
\usepackage{enumitem} % Zum Anpassen von Aufzählungen !s

\setlist[itemize]{noitemsep,topsep=0pt,leftmargin=*}
\setlist[description]{noitemsep,topsep=0pt,leftmargin=0pt,labelsep=0.5em,font=\bfseries}

\newcolumntype{Y}{>{\RaggedRight\arraybackslash}X}
\setlist[enumerate]{nosep,leftmargin=*}

\usepackage{float} % <-- DIESES PAKET HINZUFÜGEN
\usepackage[
    colorlinks=true,
    linkcolor=blue,
    citecolor=blue,
    urlcolor=blue
]{hyperref} % Für klickbare Links und Referenzen
\usepackage{orcidlink} % Für das ORCID-Icon

\usepackage[a4paper, total={6.5in, 9in}]{geometry} % Seitenränder anpassen

\title{\textbf{Countermind: A Multi-Layered Security Architecture for Large Language Models}}
\author{Dominik Schwarz\\
\small Independent Researcher\\
\small \texttt{dominikschwarz@acm.org}
}
\date{\today}

\begin{document}

\maketitle
\vspace{-2em}

\begin{abstract}
The security of Large Language Model (LLM) applications is fundamentally challenged by \enquote{form-first} attacks like prompt injection and jailbreaking, where malicious instructions are embedded within user inputs. Conventional defenses, which rely on post hoc output filtering, are often brittle and fail to address the root cause: the model's inability to distinguish trusted instructions from untrusted data \cite{AguileraMartinez2025}. This paper proposes Countermind, a multi-layered security architecture intended to shift defenses from a reactive, post hoc posture to a proactive, pre-inference, and intra-inference enforcement model. The architecture proposes a fortified perimeter designed to structurally validate and transform all inputs, and an internal governance mechanism intended to constrain the model's semantic processing pathways before an output is generated. The primary contributions of this work are conceptual designs for: (1) A \textbf{Semantic Boundary Logic (SBL)} with a mandatory, time-coupled \textbf{Text Crypter} intended to reduce the plaintext prompt injection attack surface, provided all ingestion paths are enforced. (2) A \textbf{Parameter-Space Restriction (PSR)} mechanism, leveraging principles from representation engineering \cite{Zou2023a}, to dynamically control the LLM's access to internal semantic clusters, with the goal of mitigating semantic drift and dangerous emergent behaviors. (3) A \textbf{Secure, Self-Regulating Core} that uses an OODA loop and a learning security module to adapt its defenses based on an immutable audit log. (4) A \textbf{Multimodal Input Sandbox} and \textbf{Context-Defense} mechanisms to address threats from non-textual data and long-term semantic poisoning. This paper outlines an evaluation plan designed to quantify the proposed architecture's effectiveness in reducing the Attack Success Rate (ASR) for form-first attacks and to measure its potential latency overhead.
\end{abstract}

\vspace{1em}

%========= SEKTIONEN  =========

\section{Introduction}
\label{sec:introduction}

\subsection{The Evolving Threat Landscape of the LLM Ecosystem}
\label{sec:intro_threat_landscape}

The rapid integration of Large Language Models (LLMs) into a vast array of applications, from consumer-facing chatbots to autonomous agents capable of executing complex tasks, marks a paradigm shift in software development and human-computer interaction \cite{Rossi2024}. As these models evolve into multimodal systems and intelligent agents, their capabilities expand, and so does their attack surface \cite{Xu2025}. The core security challenge stems from a fundamental design characteristic of transformer-based architectures, which is the lack of a clear distinction between trusted system instructions and untrusted user-provided data. This ambiguity is the foundational vulnerability exploited by a class of attacks known as prompt injection, where an adversary crafts input that manipulates the LLM into deviating from its intended behavior \cite{Liu2024a}.

The threat landscape is evolving at a pace that outstrips traditional security measures. Initial attacks involved simple, handcrafted "jailbreak" prompts designed to bypass safety alignments \cite{Shang2025a}. However, the field has rapidly advanced to include automated and universal prompt injection attacks that can be generated using gradient-based methods, requiring minimal training data to achieve high efficacy \cite{Chiu2025}. Adversaries are increasingly sophisticated, employing multimodal attack vectors where malicious instructions are hidden within images or other non-textual data to bypass text-based filters \cite{Cheng2024}. The proliferation of LLM-powered agents, which can interact with external tools and APIs, introduces further risks, including indirect prompt injection, privilege escalation, and the potential for malicious instructions to propagate through a network of agents in a manner akin to a computer virus \cite{Kong2025}. This escalating complexity necessitates a new security paradigm that addresses these threats at an architectural level.

\subsection{The Fallacy of Post hoc Filtering and the Rise of Form-First Attacks}
\label{sec:intro_fallacy_filtering}

The dominant security paradigm for LLMs has been largely reactive, focusing on post hoc filtering of model outputs. This approach, which includes techniques like input sanitization, keyword filtering, and even sophisticated LLM-based guardrails, treats the core model as an untrusted black box that must be contained \cite{LiFung2025}. This methodology is fundamentally flawed because it attempts to address the symptoms, such as harmful outputs, rather than the root cause. It places defenders in a perpetual \enquote{cat-and-mouse game} where they must constantly react to new attack techniques, a position of structural disadvantage \cite{Zhang2025}. Once a maliciously crafted input has been processed by the LLM, the model's internal state may already be compromised, making reliable output filtering exceedingly difficult \cite{Chen2025SafetyConstraints}.

This paper posits that the most potent threats are "form-first" attacks, where the attack's success is contingent on the structure, encoding, or format of the input, rather than purely its semantic content. These attacks target the model's processing logic at a pre-semantic level. The core argument of this work is that the LLM security problem should be reframed from a content moderation challenge to an architectural design challenge. Current approaches attempt to build a "cage" around an insecure process. A more robust solution requires redesigning the "operating system" in which the LLM functions to be secure by design. This involves creating a structural and verifiable separation between trusted instructions and untrusted data payloads, a solution that cannot be achieved through simple content filtering alone.

\subsection{The Countermind Approach: An Overview}
\label{sec:intro_countermind_overview}

In response to these challenges, we introduce Countermind, a conceptual security architecture that embodies a paradigm shift from reactive filtering to proactive, structural defense. Countermind is not designed as a simple wrapper or filter, but rather as a deeply integrated, multi-layered system that governs all interactions with a core LLM. The architecture is analogous to a medieval castle, which is not defended by a single wall but by a series of concentric, specialized defense rings. Each layer performs a distinct validation and control function, ensuring that threats are neutralized at the earliest possible stage.

The Countermind architecture is built upon four foundational pillars:
\begin{enumerate}[nosep,leftmargin=*]
  \item \textbf{Semantic Boundary Logic (SBL).} An enforced API perimeter that decomposes, validates, and authenticates incoming requests before processing, with the aim of reducing plaintext prompt exposure.
  \item \textbf{Parameter-Space Restriction (PSR).} An intra-inference control that modulates access to internal semantic clusters during decoding—via projection/masking of activations—with the goal of limiting semantic drift and unsafe behaviors.
  \item \textbf{Secure, Self-Regulating Core.} A governance and learning component that encodes high-level policies, maintains an append-only, tamper-evident audit log, and adapts its configuration in response to observed signals.
  \item \textbf{Multimodal and Contextual Defenses.} Modules for non-text modalities and long-horizon context (e.g., RAG), intended to mitigate risks from embedded instructions and semantic poisoning.
\end{enumerate}

\subsection{Contributions}
\label{sec:intro_contributions}

This paper makes the following primary contributions to the field of LLM security:
\begin{itemize}
    \item A novel fortified API perimeter (\textbf{SBL}) that deconstructs requests into Origin, Metadata, and Payload (OMP) and enforces structural integrity via a mandatory, time-coupled cryptographic payload wrapper (\textbf{Text Crypter}), \emph{is intended to reduce} the attack surface for conventional prompt injection.
    \item To the author's knowledge, one of the first conceptual applications of activation-steering principles for proactive security governance (\textbf{Parameter-Space Restriction}), moving beyond post hoc safety alignment...
    \item An architecture for a self-regulating and resilient security core that integrates constitutional principles, an immutable audit log, and an adaptive OODA loop to learn from attacks and dynamically update its own defensive posture.
    \item A holistic defense framework that extends protection to multimodal inputs (via a dedicated sandbox) and defends against long-term manipulation through advanced context management techniques (\textbf{Semantic Zoning, CDS, VKV}).
\end{itemize}
\section{Related Work}
\label{sec:related_work}

The Countermind architecture builds upon and extends several established and emerging areas of LLM security research. A comprehensive review of the current landscape provides context for its novel contributions \cite{Xu2025}.

\paragraph{Prompt Injection and Jailbreak Defenses:} A significant body of work focuses on defending against prompt injection and jailbreaking \cite{Liu2024a}. Common strategies include input validation and sanitization, context-aware filtering, and output encoding \cite{Yi2024}. Some approaches use defensive prompting or add special tokens to help the model distinguish instructions from data \cite{Schulhoff2025}. However, these test-time defenses are often less effective than training-time modifications \cite{Wehner2025}. Countermind's \textbf{SBL} and \textbf{Text Crypter} offer a fundamentally different, proactive defense that is intended to reduce the attack surface, provided all ingestion paths are enforced.

\paragraph{LLM Guardrails and Output Filtering:} The use of "guardrails" to filter LLM inputs and outputs is a common industry practice. These systems act as external safety layers, blocking prompts or responses that violate predefined policies. While useful, they represent a post hoc approach that Countermind aims to overcome. The reliance on external filters can be brittle and often fails to address the root cause of harmful generation \cite{Chen2025SafetyConstraints}. Countermind's \textbf{Parameter-Space Restriction (PSR)} moves this control from a reactive output check to a proactive, intra-inference mechanism that constrains the model's internal generative process.

\paragraph{Representation Engineering and Activation Steering:} PSR is grounded in the emerging field of Representation Engineering (RepE), which seeks to understand and control the internal representations of neural networks \cite{Zou2023a}. A key technique within RepE is activation steering, where vectors are added to a model's activations during inference to guide its behavior \cite{Turner2023}. While most applications of activation steering focus on influencing stylistic or behavioral traits (like honesty or reducing bias) \cite{Lee2024PRCAS}, Countermind's PSR is one of the first conceptual applications of this principle as a hard security gate, preventing the model from accessing entire semantic domains based on a security policy.

\paragraph{Constitutional AI and RLHF:} The dominant paradigms for model alignment are Reinforcement Learning from Human Feedback (RLHF) and Constitutional AI (CAI), which use human or AI-generated feedback to fine-tune models toward safer behavior \cite{Bai2022}. These methods modify the model's weights through training. Countermind's \textbf{Secure, Self-Regulating Core} complements these approaches by providing an architectural enforcement mechanism. Instead of relying solely on the model's learned behavior, it enforces immutable principles at runtime, offering a more resilient defense against novel attacks that bypass training-based alignments.

\paragraph{Multimodal Security:} As LLMs become multimodal, they introduce new attack surfaces where malicious content can be hidden in images, audio, or video \cite{Cheng2024}. The research in this area is still nascent. Countermind's \textbf{Multimodal Input Sandbox} provides a dedicated, pre-processing pipeline to analyze and neutralize such threats before they reach the core model, addressing a critical and growing vulnerability \cite{Chen2025}.

\paragraph{Threat Modeling for LLMs:} Established cybersecurity frameworks like STRIDE are being adapted to the unique challenges of AI and LLM systems \cite{Wu2024}. The Countermind architecture is designed with such a structured threat model in mind, providing specific countermeasures for threats like tampering, information disclosure, and elevation of privilege within the context of an LLM-powered application.
\section{Threat Model \& Assumptions}
\label{sec:threat_model}

To formally define the security context for Countermind, a threat model is established based on standard methodologies like STRIDE, adapted for the unique characteristics of LLM-powered systems \cite{Wu2024}. The model assumes a sophisticated adversary with black-box access to the system's API.

% --- Tabelle ---
\begin{tabularx}{\textwidth}{@{} l Y @{}}
\caption{Threat model based on the STRIDE methodology.}
\label{tab:threat_model}\\
\toprule
\textbf{Category} & \textbf{Description}\\
\midrule
\endfirsthead

\multicolumn{2}{l}{\tablename~\thetable\ — \textit{Continued from previous page}}\\
\toprule
\textbf{Category} & \textbf{Description}\\
\midrule
\endhead

\midrule
\multicolumn{2}{r}{\textit{Continued on next page}}\\
\endfoot

\bottomrule
\endlastfoot

\textbf{Attacker Goals} &
\begin{enumerate}
  \item \textbf{Policy bypass (jailbreak):} Coerce the LLM to generate harmful or restricted content.
  \item \textbf{Instruction hijacking:} Override system instructions to execute unauthorised functions (e.g., tool use).
  \item \textbf{Information disclosure:} Extract sensitive data from session, RAG context, or model memory.
  \item \textbf{Denial of service:} Induce cost/latency spikes via resource-intensive operations.
\end{enumerate}\\

\textbf{Attacker Capabilities} &
\begin{enumerate}
  \item \textbf{Input control:} Full control over prompts and multimodal inputs submitted to the API.
  \item \textbf{External resource control:} Ability to host malicious content the LLM is instructed to process.
  \item \textbf{Black-box access:} Interact via public API; no access to internals, weights, or secret keys.
\end{enumerate}\\

\textbf{Attack Channels} &
\begin{enumerate}
  \item \textbf{Direct API interaction:} Crafted text/image/audio or other formats.
  \item \textbf{Indirect ingestion:} Poisoning websites/documents processed via RAG or tools.
  \item \textbf{Supply chain:} Compromising plugins or pre-trained components.
\end{enumerate}\\

\textbf{Trust Roots \& Assumptions} &
\begin{enumerate}
  \item \textbf{Secure cryptography:} Secrets and primitives (e.g., Text Crypter) are protected from the adversary.
  \item \textbf{Imperfect LLM:} Standard aligned model; not inherently robust to sophisticated attacks.
  \item \textbf{Trusted computing base:} Countermind components, OS, and HW are uncompromised.
  \item \textbf{Ingestion enforcement:} The sandbox guarantee holds \emph{assuming} all request paths terminate at the Sandbox and no privileged side-channels exist.
\end{enumerate}\\

\textbf{Out-of-scope Threats} &
\begin{enumerate}
  \item \textbf{Physical attacks} on hosting infrastructure.
  \item \textbf{Insider threats} by privileged users with access to core components/keys.
  \item \textbf{Network-layer DoS} (large-scale DDoS outside the LLM compute path).
\end{enumerate}\\
\end{tabularx}

\subsection{State of the Art in LLM Attacks: Prompt Injection, Jailbreaking, and Agentic Risks}
\label{sec:threat_model_sota}

A comprehensive understanding of the threat landscape is essential for designing robust defenses. LLM attacks have evolved significantly, targeting various aspects of the model's lifecycle and deployment context.

\paragraph{Prompt Injection:} This is a class of vulnerabilities where an attacker manipulates an LLM's inputs to cause unintended behavior. These attacks are broadly categorized into two types.

\textbf{Direct prompt injection}, often referred to as \textbf{jailbreaking}, involves crafting prompts that explicitly instruct the model to override its safety guidelines or system instructions \cite{Liu2024a}. Techniques include role-playing scenarios, instruction bypassing, and obfuscation using encodings or complex language \cite{Yi2024}.

\textbf{Indirect prompt injection} occurs when an LLM processes data from an external, attacker-controlled source, such as a webpage or a document. The malicious prompt is embedded within this external data and is executed by the LLM with the privileges of the current session, potentially leading to data exfiltration or unauthorized actions \cite{Yang2024a}.

\paragraph{Jailbreaking:} As a specific form of prompt injection, jailbreaking aims to circumvent the safety alignment measures implemented during model training, such as Reinforcement Learning from Human Feedback (RLHF) \cite{Shang2025a}. The field has progressed from manually crafted prompts shared in online communities to sophisticated, automated attack generation methods. Optimization-based techniques, such as the Greedy Coordinate Gradient (GCG) attack, use gradient information to automatically find adversarial suffixes that are highly effective at jailbreaking models \cite{Zou2023b}. Comprehensive surveys have categorized the vast landscape of jailbreaking techniques, highlighting their increasing diversity and effectiveness against even state-of-the-art models \cite{Rossi2024}.

\paragraph{Agentic Risks:} The emergence of LLM-powered agents, which can use tools, maintain memory, and interact with other systems, has significantly expanded the threat surface \cite{Xu2025}. An agent's ability to call external APIs can be exploited to turn a successful prompt injection into a more severe vulnerability, such as Remote Code Execution (RCE), Server-Side Request Forgery (SSRF), or SQL Injection. Furthermore, in multi-agent systems, a compromised agent can spread malicious instructions to other agents, a phenomenon termed "prompt infection," which can lead to cascading failures or coordinated malicious behavior across the system \cite{Kong2025}. The unique interaction patterns of mobile LLM agents, which operate with elevated system privileges and interact with GUI elements, introduce another set of distinct attack surfaces, including UI manipulation and deeplink forgery \cite{Wu2025b}.

\paragraph{Multimodal and Structural Attacks:} Adversaries are increasingly exploiting non-textual modalities. Typographic attacks embed adversarial text within images, which can be imperceptible to humans but are processed by the LLM's vision encoder, bypassing text-based safety filters \cite{Cheng2024}. Similarly, attacks can be encoded within structured data formats like JSON or XML, manipulating the model's parsing and interpretation logic to execute hidden commands.

\subsection{Core Evaluation Metrics}
\label{sec:threat_model_metrics}

To quantitatively assess the effectiveness and practicality of LLM security systems, a set of standardized metrics is essential. The evaluation of Countermind is based on the following core metrics:

\begin{itemize}
    \item \textbf{Attack Success Rate (ASR):} This is the primary metric for security effectiveness, defined as the percentage of adversarial inputs that successfully elicit the intended malicious behavior from the target system. A successful attack could be a jailbreak that generates harmful content, an injection that triggers an unauthorized tool call, or any other outcome aligned with the adversary's goals. The measurement of ASR can be nuanced, considering not just binary success but also the severity of the violation and the diversity of successful attacks \cite{Freenor2025}.
    \item \textbf{Abstention Rate:} This metric measures the frequency with which the model correctly refuses to provide an answer to an unsafe, inappropriate, or unanswerable query. High abstention on harmful prompts is a key indicator of a robust safety system. It is crucial to distinguish a safe refusal from an incorrect but harmless response, as the former demonstrates proper safety alignment while the latter may indicate a failure of comprehension \cite{Wen2024}.
    \item \textbf{False-Block Rate (FBR):} Also known as the False Positive Rate, this metric quantifies the percentage of benign, legitimate user prompts that are incorrectly blocked or filtered by the security system. The FBR is a critical measure of the system's usability and utility. An overly aggressive security system with a high FBR may render the LLM application unusable for legitimate tasks, highlighting the inherent trade-off between security and functionality.
    \item \textbf{Latency Overhead:} This metric measures the additional processing time introduced by the security mechanisms compared to a baseline, undefended system. For real-time and interactive applications, minimizing latency is a critical non-functional requirement. High latency overhead can severely degrade the user experience, making it a key consideration in the practical deployment of any security architecture \cite{Li2025b}.
\end{itemize}
\section{System Overview}
\label{sec:system_overview}

\subsection{Architectural Blueprint and Trust Boundaries}
\label{sec:overview_blueprint}

The Countermind architecture is designed as a comprehensive, proactive security intermediary that sits between the end-user and the core LLM. It is not a monolithic application but a collection of specialized, interconnected modules, each operating within a clearly defined trust boundary. In security architecture, a trust boundary is a logical perimeter that separates trusted components from untrusted ones \cite{Shang2025b}. Any data crossing a trust boundary must be subject to validation and authentication.

Countermind establishes several key trust boundaries:
\begin{enumerate}
    \item \textbf{The External Boundary:} This is the primary boundary between the untrusted external world, including user inputs and external data sources, and the first component of the Countermind system, the \textbf{SBL}. All incoming data is considered hostile by default until it is rigorously validated.
    \item \textbf{The Internal Component Boundaries:} Each major component within Countermind, such as the \textbf{SBL}, \textbf{PSR}, Secure Core, and Sandbox, operates in its own logical space. The interfaces between these components are themselves trust boundaries, ensuring that a potential compromise in one module does not automatically grant access to others. For example, the output of the \textbf{SBL} is validated before being passed to the PSR module.
    \item \textbf{The Core Model Boundary:} This is the final boundary between the Countermind governance framework and the core LLM itself. The PSR acts as the gatekeeper at this boundary, controlling precisely what semantic information is allowed to influence the model's generative process.
\end{enumerate}

This structured approach, visualized in the system's block diagram, ensures that security is enforced at multiple stages of the request lifecycle, adhering to the principle of defense-in-depth.

\begin{figure}[H]
    \centering
    % Wir binden jetzt die PNG-Datei ein
    \includegraphics[width=0.9\textwidth]{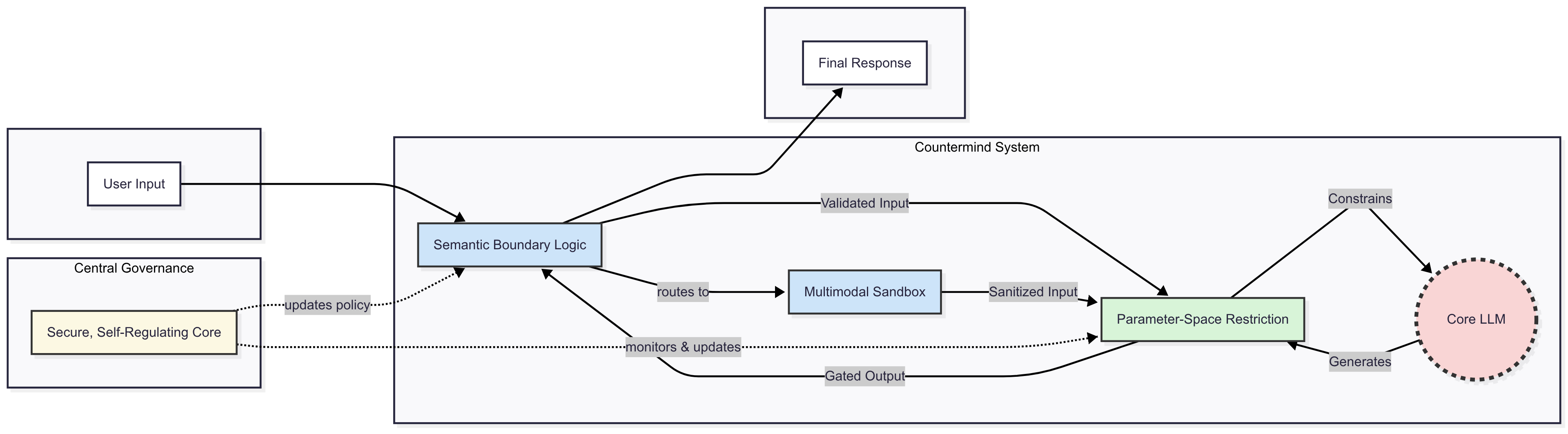}
    \caption{Multi-Layer Overview of the Countermind Architecture.}
    \label{fig:architecture_overview}
\end{figure}

\subsection{Core Components and Signal Flow}
\label{sec:overview_components}

The architecture consists of four primary components, each responsible for a distinct aspect of security enforcement.
\begin{itemize}
    \item \textbf{SBL} This is the outermost defense layer, acting as the system's fortified API gateway. It is responsible for the initial triage, syntactic validation, cryptographic verification, and semantic routing of all incoming requests before they are permitted to proceed deeper into the system.
    \item \textbf{Parameter-Space Restriction (PSR):} This is an innovative intra-inference control mechanism. It functions as a \enquote{semantic shield} by dynamically gating the LLM's access to its own internal activation space based on a granular policy. This prevents the model from generating outputs based on prohibited or irrelevant concepts.
    \item \textbf{Secure, Self-Regulating Core:} This is the central governance and intelligence hub of the system. It comprises a Semantic Trust Core, which enforces immutable "constitutional" principles, and a Learning Security Core, which monitors system behavior, learns from attack patterns, and adaptively updates the system's defensive posture.
    \item \textbf{Multimodal Input Sandbox:} This is a specialized gateway, enforced by gating and mandatory routing, where attempted bypasses are blocked and audited. It is designed to analyze, sanitize, and neutralize threats embedded in non-textual inputs, such as images, videos, and audio files, before they can influence the core LLM.
\end{itemize}

The typical signal flow for a user request is as follows: A request from a user first enters the \textbf{SBL}. If the request contains multimodal data, it is routed to the \textbf{Multimodal Input Sandbox} for analysis. Once the input is validated and sanitized, it is passed to the \textbf{Parameter-Space Restriction (PSR)} module, which determines the allowed semantic clusters for processing. The request, along with the PSR policy, is then sent to the \textbf{Core LLM}. The LLM's generation process is constrained by the PSR policy. The generated output is again gated by the PSR before being passed back through the \textbf{SBL} to the user. Throughout this process, the \textbf{Secure, Self-Regulating Core} monitors all interactions, logs them immutably, and uses this data to update the policies of both the \textbf{SBL} and the PSR module via a continuous feedback loop.

\subsection{Design Goals and Non-Goals}
\label{sec:overview_goals}

To clarify the scope and intent of the Countermind architecture, it is essential to define its explicit design goals and non-goals.

\paragraph{Design Goals:}
\begin{itemize}
    \item \textbf{Proactive Threat Neutralization:} The primary goal is to detect and block attacks at the earliest possible stage of the request lifecycle, ideally before the core LLM is ever invoked.
    \item \textbf{Structural Security:} Defenses are based on verifiable and deterministic properties of the input, such as its cryptographic integrity, structure, and format, rather than relying solely on the fallible and often brittle semantic analysis of natural language.
    \item \textbf{Defense-in-Depth:} The architecture is designed such that the failure or bypass of a single defensive layer does not lead to a full system compromise. Redundant and overlapping controls provide resilience.
    \item \textbf{Adaptability:} The system is designed to be dynamic, enabling the Learning Security Core to analyze observed attack patterns and automatically update its own defensive policies over time.
    \item \textbf{Interpretability and Auditability:} Security decisions made by the system, particularly those by the PSR module, are intended to be traceable to explicit policies. The immutable audit log ensures that all system actions are transparent and available for forensic analysis.
\end{itemize}

\paragraph{Non-Goals:}
\begin{itemize}
    \item \textbf{Perfect LLM Alignment:} Countermind does not aim to "fix" or fundamentally alter the weights of the core LLM. It operates on the assumption that the LLM is an imperfect and potentially vulnerable component, and its goal is to govern the \textit{use} of this component securely.
    \item \textbf{Elimination of All Harmful Content:} The objective is not to create a completely "censored" model that refuses all sensitive topics. Rather, the goal is to prevent malicious \textit{hijacking} of the model's capabilities and to enforce a configurable, policy-driven security posture.
    \item \textbf{Zero Performance Overhead:} The architecture explicitly acknowledges that robust security measures incur a performance cost. The goal is not to eliminate this overhead but to make it quantifiable, manageable, and a deliberate trade-off in the system's design.
\end{itemize}
\section{Semantic Boundary Logic: A Fortified API Perimeter}
\label{sec:semantic_boundary_logic}

The \textbf{SBL} is the first and most critical line of defense in the Countermind architecture. It transforms the traditional concept of an API endpoint from a simple data receiver into an intelligent, multi-stage validation and control system. Its purpose is to filter and validate the \textit{intention} and \textit{structure} of an input long before it reaches the core model.

\subsection{The Origin-Metadata-Payload (OMP) Decomposition}
\label{sec:sbl_omp}

The foundational principle of the \textbf{SBL} is the systematic deconstruction of every incoming request into three distinct, analyzable components. This OMP decomposition allows for a granular, multi-faceted risk assessment.
\begin{itemize}
    \item \textbf{Origin:} This component represents the source of the request. The system validates the origin against a registry of known and trusted sources, such as a specific version of a mobile application, a registered third-party plugin, or an internal development client. The origin's identity, session integrity, and historical behavior are the first factors in the trust assessment.
    \item \textbf{Metadata:} This component provides structured information about the nature of the data being transmitted. It declares the payload type (e.g., \texttt{INPUT\_PLAINTEXT}, \texttt{INPUT\_PICTURE}, \texttt{INPUT\_VIDEO}) and the expected format. This declaration allows the system to apply the correct validation rules and routing logic without having to first infer the data type from the raw content.
    \item \textbf{Payload:} This is the core content of the request. Crucially, the \textbf{SBL} enforces strict rules not just on the semantic content but on the \textit{form} of the payload. For text-based inputs, the payload is required to be encapsulated in a self-validating, authenticated encoding generated by the Text Crypter. This includes an embedded HMAC-SHA256 to prove its integrity and authenticity.
\end{itemize}

\subsection{Layered Defense: Syntactic Gate, Intent-Based Router, and Semantic Filter}
\label{sec:sbl_layered_defense}

Building on the OMP structure, a three-layer defense progressively deepens its analysis of the request.
\begin{itemize}
    \item \textbf{Layer 1 – Syntactic Gate:} This is the first hard barrier, acting as an unyielding "byte-gate." It performs purely structural validation without any semantic interpretation. Its functions include:
    \begin{itemize}
        \item \textbf{Strict Character Allowlist:} Only characters from a predefined, minimal set (e.g., the Base62 output of the Text Crypter) are accepted. This immediately blocks a wide range of obfuscation and injection attacks that rely on complex Unicode characters, zero-width spaces, or other encoding tricks.
        \item \textbf{Cryptographic Integrity Check:} It validates the embedded HMAC-SHA256 of the payload using a shared secret key. Any payload that has been tampered with in transit or was not generated by a legitimate client fails this check. The use of HMAC-SHA256 provides strong guarantees of both frame integrity and message authenticity, while a simpler CRC might only be used for faster, non-cryptographic frame integrity checks.
    \end{itemize}
    The vast majority of automated, fuzzing-based, and simple injection attacks are defeated at this fundamental layer.

    \item \textbf{Layer 2 – Intent-Based Routing:} Once a request passes the syntactic gate, this layer performs an initial semantic classification to determine its intended \textit{function}, not its detailed content. Based on a dynamic "Base Table," it routes the request to the appropriate downstream handler. For example, a request classified as \texttt{CodeFragment.Analysis} is routed to an isolated sandbox environment, while a request classified as \texttt{System.Config.Modification} is blocked or routed to a high-security admin interface requiring multi-factor authentication. A standard conversational query is routed towards the core LLM via the PSR. This routing table is not static; it can be dynamically updated based on context, such as time of day or the user's current trust score. Policy updates are versioned, and the version hash is included in the audit log, ensuring that every runtime decision is deterministically reproducible from the combination of (\texttt{policy\_version}, \texttt{inputs}).

    \item \textbf{Layer 3 – Semantic Filter \& Trust Engine:} This is the deepest layer of the perimeter, performing a more nuanced analysis of user intent and behavior. It looks for complex patterns indicative of malicious intent, such as sudden and illogical topic changes, hidden instructions within seemingly benign text, or attempts at semantic obfuscation. This layer maintains a session-based \textbf{Trust Score} for each user or origin. The score is calculated using a fractal logic where consistent, safe interactions slowly increase the score, while a single suspicious request can cause a radical and significant decrease. If the Trust Score falls below a predefined threshold, the \textbf{Soft-Lock Engine} is activated. This engine intercepts the final bot response and replaces it with a graceful degradation response, effectively degrading the service for the suspicious user. The governance of this mechanism, including threshold calibration, automatic unfreezing policies, and pathways for human review of false positives, is detailed in the Ethics Statement.
\end{itemize}

\subsection{The Text Crypter: Authenticated Envelope (No Custom Crypto)}
\label{sec:sbl_text_crypter}

A cornerstone of the \textbf{SBL} is the mandatory use of an \emph{authenticated, time-bound envelope} for all text-based payloads. We rely on \emph{standard message authentication} (e.g., HMAC-SHA-256; JWS/PASETO-style MAC) with nonce and TTL. This is \emph{not encryption and not custom cryptography}.

\paragraph{Security intent.}
No unverified plaintext enters the core system. The boundary shifts to the client: semantic intent is framed and authenticated before transmission.

\paragraph{Mechanism (three stages).}
\begin{enumerate}
  \item \textbf{Canonical envelope.} Build a canonical JSON-like envelope $E$ with metadata and the \emph{original UTF-8 payload as-is}:
  \begin{itemize}
    \item \texttt{alg} (e.g., \texttt{HMAC-SHA-256}), optional \texttt{kid}
    \item \texttt{nonce} (unique), \texttt{iat} (issued-at), \texttt{exp} (expiry = iat+TTL)
    \item \texttt{payload\_b64url} (base64url of the raw UTF-8 payload) and \texttt{payload\_sha256} (integrity hint)
  \end{itemize}

  \item \textbf{Authentication \& anti-replay.} Compute \texttt{mac = HMAC(k, canonical(E without mac))}. The server keeps an anti-replay cache keyed by (\texttt{nonce}, \texttt{kid}) and enforces \texttt{iat}/\texttt{exp}. Keys rotate periodically.

  \item \textbf{Deterministic serialization (transport).} Envelope \emph{metadata} uses a strict Base62/URL-safe alphabet; the \emph{payload remains UTF-8} and is only base64url-encoded. \emph{No dynamic alphabet remapping of user content.} This is transport framing, not secrecy.
\end{enumerate}

\paragraph{I18N/UX.}
The allowlist applies to the \emph{envelope fields} (control plane) only. The user \emph{payload} remains full UTF-8/Unicode and is MAC-protected as-is; international text and emojis are preserved.

\paragraph{Optional micro-integrity (defense-in-depth).}
Per-segment tags (e.g., hashes/HMACs for words/phrases) may be included as non-authoritative hints inside the payload and are \emph{not} required for security decisions; the authoritative check is the envelope MAC.

\paragraph{Server-side verification.}
On receipt: (1) validate \texttt{mac}, (2) check \texttt{nonce}/anti-replay and time window, (3) decode \texttt{payload\_b64url} back to UTF-8, (4) optionally verify micro-tags if present. Any failure leads to immediate rejection.

\paragraph{Example (conceptual).}
\texttt{\small \{alg:"HMAC-SHA-256", kid:"k1", nonce:"…", iat:1723833600, exp:1723833660, payload\_sha256:"…", payload\_b64url:"…", mac:"…"\}}

\begin{tabularx}{\textwidth}{@{} l YYYY @{}}
  \caption{Overview of \textbf{SBL} modules.}
  \label{tab:sbl_modules}\\
  \toprule
  \textbf{Module} & \textbf{Purpose} & \textbf{Inputs} & \textbf{Outputs} & \textbf{Performance implication} \\
  \midrule
  \endfirsthead

  \multicolumn{5}{l}{\tablename~\thetable\ — \textit{Continued from previous page}}\\
  \toprule
  \textbf{Module} & \textbf{Purpose} & \textbf{Inputs} & \textbf{Outputs} & \textbf{Performance implication} \\
  \midrule
  \endhead

  \midrule
  \multicolumn{5}{r}{\textit{Continued on next page}}\\
  \endfoot

  \bottomrule
  \endlastfoot

  \textbf{Byte-Gate} &
  Enforces syntactic validity and character restrictions; blocks non-allowlisted characters and malformed frames. &
  Raw request string. &
  Verdict (Pass/Fail), reason. &
  Very low (byte-level checks; no NLP). \\
  \addlinespace

  \textbf{Text Crypter (Server)} &
  Decodes and validates time-coupled, authenticated payloads. &
  Encoded payload from Byte-Gate. &
  Decoded plaintext; integrity verdict (Pass/Fail). &
  High (cryptographic operations; deterministic). \\
  \addlinespace

  \textbf{Base Table Router} &
  Classifies request intent (metadata-aware) and routes to the appropriate handler (e.g., Core LLM, Sandbox). &
  Decoded plaintext; metadata. &
  Routing decision (e.g., \texttt{ROUTE\_TO\_SANDBOX}). &
  Medium (semantic classification). \\
  \addlinespace

  \textbf{Trust-Scaler} &
  Computes session trust from interaction history (topic shifts, inconsistencies, prior verdicts). &
  History of requests and verdicts. &
  Updated trust score (float). &
  Medium (stateful scoring). \\
  \addlinespace

  \textbf{Soft-Lock Engine} &
  Produces graceful degradation responses when trust is low. &
  Low-trust signal; original bot response. &
  Modified user-facing response. &
  Low (lightweight proxy). \\
\end{tabularx}
\section{Parameter-Space Restriction (PSR): Controlling Semantic Activation}
\label{sec:psr}

While the \textbf{SBL} fortifies the system's perimeter, the Parameter-Space Restriction (PSR) mechanism provides a powerful, second layer of defense that operates deep within the model's generative process. It represents a fundamental shift from filtering outputs to proactively controlling the internal "thought process" of the LLM.

\subsection{Motivation: Preventing Semantic Drift and Dangerous Emergence}
\label{sec:psr_motivation}

The core motivation for PSR is the acknowledgment of the "post hoc filter fallacy." Relying on filtering the final text output is insufficient because harmful or undesirable content often arises from the emergent combination of seemingly benign concepts within the model's vast, high-dimensional parameter space. This leads to two primary failure modes:
\begin{itemize}
    \item \textbf{Semantic Drift:} During the generation process, the model's internal representation of a concept can "drift" into an unsafe or irrelevant semantic domain due to probabilistic associations learned from its training data. For example, a query about "how to bake a cake" might activate internal representations related to heat, chemical reactions, and expansion, which could be semantically adjacent to concepts related to chemistry or even explosives. While the final output may be harmless, the model has traversed a risky internal pathway. The study of semantic change in linguistics provides a formal basis for understanding how word meanings can shift and evolve over time, a phenomenon mirrored within the latent space of LLMs \cite{LaMalfa2024}.
    \item \textbf{Dangerous Emergence:} LLMs can synthesize new, dangerous information by combining knowledge from disparate, individually harmless domains. For example, a model could combine its knowledge of public software libraries with its knowledge of common vulnerability patterns to generate a novel, working exploit. Post hoc filters that look for known malicious code signatures would fail to detect such a novel synthesis.
\end{itemize}
PSR is designed to prevent these failures at their source by structurally limiting the "conceptual vocabulary" the model is allowed to use for any given task.

\subsection{Mechanism: An Activation-Steering Bridge for Pre-Output Gating}
\label{sec:psr_mechanism}

The technical implementation of PSR is best understood as a security-oriented application of \textbf{Representation Engineering (RepE)}, a field that focuses on understanding and manipulating the internal representations (activations) of neural networks \cite{Zou2023a}. Specifically, PSR leverages principles from \textbf{Activation Steering}, where vectors are added to the model's activations during inference to guide its behavior \cite{Turner2023}.

However, unlike typical applications of activation steering which aim to subtly influence style, PSR uses this mechanism as a hard \textbf{gating function}. It operates as an "activation-steering bridge" that is placed logically between the model's final processing layers and its output token selection layer. The mechanism works as follows:
\begin{enumerate}
    \item A request, having been validated by the \textbf{SBL}, is mapped to a policy that defines a set of allowed ``semantic clusters.''
    \item During the LLM's forward pass, at each generation step, the PSR mechanism intervenes on the model's hidden state activations.
    \item It ensures that the activations are constrained to lie within the subspace corresponding to the allowed semantic clusters. Activations pointing towards prohibited clusters are suppressed or zeroed out. This hard gating can degrade output quality, so a ``soft gating'' approach, where $A'_{l} = \alpha \cdot A_{l} + (1-\alpha) \cdot \Pi(A_{l})$ and $\alpha$ is a data-driven coefficient, is a potential refinement to balance safety and utility.
    \item Only then is the modified activation passed to the final layer for predicting the next token.
\end{enumerate}
This approach is a form of "cognitive alignment" rather than "behavioral alignment." Instead of training the model to \textit{act} safely after the fact (as in RLHF), PSR constrains what the model is allowed to \textit{think about} in the first place. This provides a more fundamental and robust form of control, as it is resilient to novel linguistic phrasings of a malicious request. If the entire semantic cluster related to "weapon manufacturing" is disabled by the PSR policy, no amount of clever wording should be able to activate the corresponding internal representations needed to generate a harmful response. We leave precise overhead quantification to future work; preliminary profiling indicates a modest overhead for projection and a lower overhead for masking.
\newpage
\paragraph{PSR hook, operator, cost.}
Let $y \in \mathbb{R}^d$ be the residual stream at decode step $t$ (after the last $N$ transformer blocks).
\\Let $\Pi \in \mathbb{R}^{d \times k}$ have orthonormal columns spanning the allowed subspace and define the projector $P := \Pi \Pi^{\mathsf{T}}$.
\\ \\We gate decoding by
\[
  y' = \alpha\, y + (1-\alpha)\, P y , \quad \alpha \in [0,1].
\]\\
Hard gating is obtained with $\alpha=0$, while soft gating trades safety for utility via policy-driven $\alpha$.
Hook points are the residual streams of the last $N$ blocks (typically $N \in \{1,\ldots,4\}$).
The computational cost per token is $O(d\,k)$ for the projection (compute $v=\Pi^{\mathsf{T}} y$ and then $y'=\alpha y + (1-\alpha)\Pi v$).
The matrix~$\Pi$ is precomputed/frozen; no backpropagation is involved at inference.

\subsection{Governance Through Semantic Clusters and Prefix-Rights}
\label{sec:psr_governance}

The governance of the PSR mechanism is based on a simple yet powerful policy language comprising two core concepts.
\begin{itemize}
    \item \textbf{Semantic Clusters:} The model's vast parameter space is conceptually partitioned into logical, thematic domains. These are not necessarily disjoint but represent distinct areas of knowledge or capability (e.g., \texttt{Code.Python}, \texttt{History.WWII}, \texttt{Biology.Genetics}, \texttt{System.InternalAPIs}). The definition of these clusters is a critical engineering task, likely requiring a combination of automated clustering techniques on activation data and expert human oversight.
    \item \textbf{Prefix-Rights:} For each incoming query, a policy is generated that assigns specific, granular rights to one or more semantic clusters. These rights are expressed as prefixes and define the permissible operations:
    \begin{itemize}
        \item \textbf{READ:} Allows the model to access and reproduce information from the cluster. This is the most restrictive right, suitable for factual recall.
        \item \textbf{SYNTH:} Allows the model to synthesize new information and create novel combinations \textit{within} the activated cluster. This is required for creative tasks like writing a new story or generating a new recipe.
        \item \textbf{EVAL:} Allows the model to evaluate an external input (e.g., a piece of code provided by the user) in the context of the cluster's knowledge. For example, EVAL on the \texttt{Security.CodeAnalysis} cluster would allow the model to check for vulnerabilities without executing the code.
        \item \textbf{CROSS:} A highly privileged and restricted right that allows the controlled combination of information between two or more explicitly specified clusters. This is necessary for complex, multi-domain queries but is disabled by default to prevent unintended information leakage.
    \end{itemize}
    \item \textbf{Thematic Isolation \& Query Decomposition:} The default security posture of PSR is \textbf{strict thematic isolation}. No cross-talk between clusters is permitted unless explicitly authorized by a CROSS right. Complex user queries that touch upon multiple domains are first decomposed by a pre-processing step. The system then processes the request sequentially, activating the \texttt{Code.Python} cluster with SYNTH rights to generate the script structure, and then activating the \texttt{Finance.MarketData} cluster with READ rights to provide the context, all while preventing any direct, uncontrolled interaction between them. The PSR policy for RAG-sourced context (e.g., from \texttt{RAG.*} clusters) would explicitly deny SYNTH or EVAL rights, preventing the model from interpreting retrieved text as instructions.
    \item \textbf{Governance Hook:} The PSR framework provides a powerful hook for implementing a system-level "Constitution." High-level safety and ethical principles can be translated into permanent PSR rules. For example, a constitution could enforce a permanent policy that grants only READ rights to clusters deemed sensitive or completely disables all rights for clusters deemed dangerous. A concrete policy rule might look like: \texttt{deny CROSS(*, Tools.*) unless user\_role=developer\_privileged}.
\end{itemize}

\noindent\textit{Policy artifact.} A deny-by-default cluster policy (YAML) used by our prototype is provided in Appendix~\ref{app:policy}; PSR gates $\alpha$, tool access, and cross-cluster rights accordingly.

\keepXColumns
\begin{tabularx}{\textwidth}{@{} 
    >{\raggedright\arraybackslash\hsize=1.2\hsize}X  % Spalte 1: 120% der Basisbreite
    >{\raggedright\arraybackslash\hsize=1.3\hsize}X  % Spalte 2: 130% der Basisbreite
    >{\raggedright\arraybackslash\hsize=0.4\hsize}X  % Spalte 3: 40% der Basisbreite (schmal)
    >{\raggedright\arraybackslash\hsize=1.05\hsize}X % Spalte 4: 105% der Basisbreite
    >{\raggedright\arraybackslash\hsize=1.05\hsize}X % Spalte 5: 105% der Basisbreite
@{}}
    \caption{Example of PSR Policy Assignment based on User Intent.}
    \label{tab:psr_policy} \\
    \toprule
    \textbf{User Intent \newline (Example)} & \textbf{Activated \newline Cluster(s)} & \textbf{Assigned \newline Rights} & \textbf{Default \newline Policy} & \textbf{Override Rule \newline Example} \\
    \midrule
    \endfirsthead % Definiert das Ende des Kopfes für die ERSTE Seite
    
    \multicolumn{5}{l}{\tablename~\thetable\ -- \textit{Continued from previous page}} \\
    \toprule
    \textbf{User Intent \newline (Example)} & \textbf{Activated \newline Cluster(s)} & \textbf{Assigned \newline Rights} & \textbf{Default \newline Policy} & \textbf{Override Rule \newline Example} \\
    \midrule
    \endhead % Definiert den Kopf für alle FOLGESEITEN
    
    \bottomrule
    \endlastfoot % Definiert den Fuß für die LETZTE Seite

    "Explain 'Hello World' in C++" & \texttt{Code.\hspace{0pt}C++.\hspace{0pt}Reference} & READ & Deny synthesis of new, creative code. & If user is a trusted developer, elevate to SYNTH to allow code generation. \\
    \addlinespace
    
    "Write a new recipe for cheesecake" & \texttt{Kitchen.\hspace{0pt}Recipes.\hspace{0pt}Desserts} & SYNTH & Allow combination of ingredients within the cluster. Isolate from \texttt{Chemistry.\hspace{0pt}LabSafety}. & N/A \\
    \addlinespace
    
    "Analyze this code for SQL injection" & \texttt{Security.\hspace{0pt}CodeAnalysis.\hspace{0pt}SQLi} & EVAL & Deny execution of the code. Isolate from \texttt{System.\hspace{0pt}Database.\hspace{0pt}Access}. & If in a sandboxed test environment, allow CROSS with \texttt{System.\hspace{0pt}Database.\hspace{0pt}Access} for live testing. \\
    \addlinespace
    
    "Write a C++ program to bake a cake" & 1. \texttt{Code.\hspace{0pt}C++.\hspace{0pt}Reference} \newline 2. \texttt{Kitchen.\hspace{0pt}Recipes.\hspace{0pt}Desserts} & 1. SYNTH \newline 2. READ & Process sequentially. Deny CROSS between clusters to prevent semantic drift. & A highly privileged "creative agent" might be granted CROSS rights for novel combinations. \\
\end{tabularx}
\section{The Secure, Self-Regulating Core}
\label{sec:secure_core}

The Secure, Self-Regulating Core is the central intelligence and governance unit of the Countermind architecture. It ensures the system's long-term integrity, enforces foundational principles, and enables the architecture to adapt and evolve its defenses over time. It transforms the system from a static set of filters into a resilient, learning entity.

\subsection{Principles: Constitutional Enforcement and Immutable Audit Logs}
\label{sec:core_principles}

The operation of the Secure Core is grounded in several fundamental principles.
\begin{itemize}
    \item \textbf{Cluster-Scoped Modification Rights:} The architecture envisions a future state where the AI can learn and even modify its own algorithms to improve performance and security. However, this capability is strictly governed. The AI is only granted modification rights within specific, non-critical semantic clusters. Core security policies, constitutional principles, and the logic of the Secure Core itself are defined in immutable, "read-only" clusters.
    \item \textbf{Protected Core Semantic Layers:} Certain foundational knowledge domains and reasoning pathways within the LLM are designated as protected layers. These act as the AI's "genetic code," preserving its core identity and safety alignment, and are exempt from any form of algorithmic self-modification.
    \item \textbf{Immutable Audit Log:} Every significant event within the system is recorded in a cryptographically secured, tamper-evident audit log. This includes every request, every validation verdict, every PSR policy decision, every tool call, and every proposed self-modification. The principle of immutability, often implemented using techniques analogous to write-once-read-many (WORM) storage or Merkle trees, is critical. It ensures that the log is a verifiable and trustworthy record of the system's history, which is essential for forensic analysis, compliance, and as the ground truth for the system's own learning processes.
\end{itemize}

\subsection{The Semantic Trust and Learning Security Cores}
\label{sec:core_modules}

The Secure Core is composed of two primary, interconnected modules.
\begin{itemize}
    \item \textbf{Semantic Trust Core:} This module functions as the system's constitutional arbiter. It contains the encoded, immutable principles that govern the AI's behavior. Its primary role is to provide a final check and veto power over critical actions. For example, if the Learning Core proposes a change to a PSR policy, the Trust Core validates that this change does not violate any fundamental safety principles. This architectural enforcement of a constitution provides a more robust and transparent mechanism than purely training-based approaches like Constitutional AI, which rely on the model learning to follow principles through reinforcement learning \cite{Bai2022}.
    \item \textbf{Learning Security Core:} This is the adaptive engine of the architecture. It is responsible for detecting threats and evolving the system's defenses. It comprises several sub-components:
    \begin{itemize}
        \item \textbf{Delta-Monitor:} A "semantic seismograph" that continuously monitors interaction patterns and the usage of semantic clusters over time. It is designed to detect subtle, low-and-slow attacks, such as gradual context poisoning or the slow erosion of cluster boundaries.
        \item \textbf{Intent-Detector:} A deep analysis module that goes beyond the surface-level syntax of a prompt to infer the user's true intention. It compares this inferred intent against the defined purpose of the semantic clusters being requested, flagging mismatches that could indicate a "cluster-mimese" attack, where a malicious request is disguised to fit a benign category.
        \item \textbf{Asynchronous Audit:} A background process that continuously analyzes the immutable audit log. It uses pattern recognition and anomaly detection to identify novel attack signatures and trends. The findings from this audit are the primary input for adapting the system's security posture.
    \end{itemize}
\end{itemize}

\subsection{Context-Defense: Semantic Zoning and Versioned Key-Value (VKV) Context}
\label{sec:core_context_defense}

The architecture explicitly treats the LLM's conversation context as a primary vulnerability. Unstructured and unvalidated context can be exploited for long-term manipulation, a threat known as context hijacking or semantic poisoning. The ever-expanding context windows of modern LLMs increase this attack surface, creating a "needle in a haystack" problem where malicious instructions can be hidden deep within a long conversation history \cite{Balarabe2025}. To counter this, the Secure Core implements a suite of context-defense mechanisms.
\begin{itemize}
    \item \textbf{Semantic Zoning:} The conversation history is not stored as a single, monolithic block of text. Instead, it is partitioned into structured \textbf{Semantic Zones}, each with a clear purpose and its own set of access rights. For example, information from a user's casual small talk is placed in a \texttt{DIALOG.SMALLTALK} zone with very limited rights, while code snippets being analyzed are placed in a \texttt{TECH.CODE.ANALYSIS} zone. Information is not allowed to "leak" or influence processing between zones without an explicit, policy-driven authorization. This prevents information provided in one context from being inappropriately used in another.
    \item \textbf{Context-Delta Sentinel (CDS):} This is a specialized watchdog module that monitors the semantic integrity of the context over time. It identifies persistent concepts or entities within the conversation history and tracks how their internal representation and semantic weight evolve. If the CDS detects a significant and unauthorized "semantic drift," for example, a seemingly harmless term like "Admin-Panel" gradually acquiring strong semantic associations with privileged system commands through repeated, subtle user references, it raises an alarm. This can trigger a corrective action, such as resetting the semantic weight of the poisoned term to its original, safe state.
    \item \textbf{Versioned Key-Value (VKV) Context:} To provide a robust defense against the recursive effects of context poisoning, the context is managed using a versioning system, similar to git in software development. Every significant update to the context does not overwrite the old state but creates a new, versioned entry. The LLM's operations are always bound to the latest \textit{validated and authorized} version of the context. This prevents an attacker from poisoning the context and then having the model recursively act upon that poisoned state in a feedback loop. Even if a malicious version is created, it cannot become the new ground truth for the system without passing through a validation process.
\end{itemize}

\subsection{The Adaptive OODA Loop for Dynamic Security Updates}
\label{sec:core_ooda}

The entire process of threat detection and response within the Secure Core is modeled on the \textbf{Observe, Orient, Decide, Act (OODA) loop}, a framework developed for decision-making in fast-paced, adversarial environments. This enables the system to operate as an autonomous cyber defense agent for itself.
\begin{itemize}
    \item \textbf{Observe:} The Delta-Monitor, Intent-Detector, and other logging mechanisms continuously gather data from user interactions and the internal state of the system. This is the data collection phase.
    \item \textbf{Orient:} The Asynchronous Audit process analyzes the observed data, placing it in the broader context of past events (from the immutable log) and the current security policies. This phase is where raw data is turned into actionable intelligence, identifying potential threats or anomalies.
    \item \textbf{Decide:} Based on the orientation, the Learning Security Core determines the optimal course of action. This could range from a minor adjustment, like lowering a user's Trust Score, to a significant policy change, like creating a new rule for the PSR module to block a novel attack pattern. This decision is then validated by the Semantic Trust Core against the system's constitution.
    \item \textbf{Act:} The system executes the decision. This could involve dynamically pushing an updated rule set to the PSR, activating the Soft-Lock engine for a specific user, or flagging a new pattern for human review. This action alters the system's defensive posture, and the loop begins again.
\end{itemize}
This adaptive OODA loop allows Countermind to evolve its defenses without requiring constant manual intervention or full model retraining, making it resilient against a changing threat landscape.

\begin{figure}[H]
    \centering
    \includegraphics[width=0.9\textwidth]{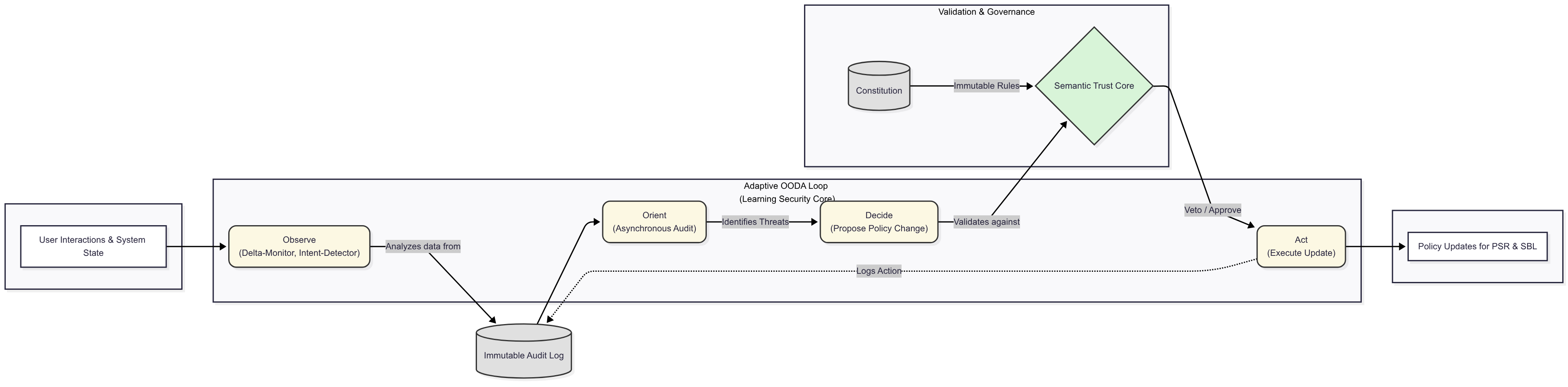}
    \caption{Trust/Learning Cores – Data Paths \& Feedback Loops.}
    \label{fig:cores_feedback_loop}
\end{figure}
% sections/08_multimodal_sandbox.tex

\section{Multimodal Sandbox}
\label{sec:multimodal-sandbox}

To address threats that bypass text-only defenses, Countermind incorporates a dedicated sandbox for all non-textual inputs. This is enforced by mandatory routing, where any attempted bypass is blocked and audited. This component is critical, as adversaries increasingly hide malicious prompts and content within images, videos, and audio files \cite{Cheng2024}. The primary risk is that a text-based prompt may appear benign, while the accompanying multimodal input contains the true malicious payload, such as an image with embedded text instructions or a video intended for deepfake generation.

\subsection{Pipeline Overview and Mandatory Routing}
\label{sec:sandbox-pipeline}

The sandbox operates as an isolated, multi-stage analysis pipeline that every multimodal input must pass through before it can be processed by the core LLM. The system is designed to ensure all non-textual data is gated through this process without exception, providing a single point of control and auditing.

\subsection{Preprocessors}
\label{sec:sandbox-preprocessors}

The first stage of the pipeline involves disassembling and analyzing the raw input data to extract meaningful features for security analysis.

\subsubsection{Video}
\label{sec:sandbox-video}

\paragraph{Disassembly} Videos are decomposed into individual frames using tools like \textbf{FFmpeg} to enable granular analysis of the visual content, frame by frame.

\subsubsection{Images}
\label{sec:sandbox-images}

\paragraph{Perceptual Hashing (pHash)} A unique "fingerprint" is generated for each image or video frame. This hash is compared against a database of known illicit or harmful content (e.g., CSAM). While pHash is robust to minor modifications like resizing or compression, it can be vulnerable to more sophisticated adversarial manipulations \cite{Yang2025}.

\paragraph{Face-ID Matching} Using libraries like InsightFace or \textbf{MediaPipe}, the system performs precise detection of real, identifiable human faces. This is a critical step to prevent the misuse of the system for generating non-consensual deepfakes.

\paragraph{Nudity Classification} A specialized neural network classifier (e.g., NudeNet) assesses each frame for the presence of explicit or implied nudity.

\subsubsection{Audio}
\label{sec:sandbox-audio}

\paragraph{Threat Vector} Audio inputs represent a significant vector for prompt injection, as malicious instructions can be delivered vocally, bypassing filters that analyze only typed text. The primary risks include spoken jailbreak commands, social engineering attempts, or instructions hidden within seemingly benign audio recordings.

\paragraph{Analysis Pipeline} The audio analysis pipeline is a multi-stage process designed to convert spoken language into a sanitized, analyzable text format.
\begin{itemize}
    \item \textbf{Voice Activity Detection (VAD):} As a crucial preprocessing step, a VAD module first analyzes the audio stream to isolate segments containing human speech. This allows the system to discard silence and non-relevant background noise, optimizing efficiency by ensuring that only meaningful data is passed to the more resource-intensive transcription stage.
    \item \textbf{Automatic Speech Recognition (ASR):} Each speech segment identified by the VAD is then processed by a high-accuracy ASR model. This model transcribes the spoken words into a raw text string. The choice of ASR model is critical, as transcription errors could potentially obscure or alter the intent of a malicious prompt.
    \item \textbf{Text-Based Guardrail Integration:} The transcribed text is not trusted by default. It is immediately treated as a standard user input and routed back to the main security pipeline, beginning with the \textbf{SBL}. This ensures that spoken prompts undergo the same rigorous analysis for injection, policy violations, and harmful content as any typed input, thereby unifying the defense strategy across modalities.
\end{itemize}

\subsubsection{Documents}
\label{sec:sandbox-documents}

\paragraph{Threat Vector} Documents from untrusted sources (e.g., PDFs, Word files) are a classic vector for indirect prompt injection. An adversary can embed malicious text in various ways: hidden in metadata, written in a color invisible to the human eye (e.g., white text on a white background), or embedded within images inside the document. Furthermore, formats like DOCX or PDF can contain malicious macros or scripts.

\paragraph{Analysis Pipeline} The document sandbox deconstructs files to extract and analyze all potential instruction-carrying content.
\begin{itemize}
    \item \textbf{MIME Type and Magic Number Validation:} The analysis begins with a strict verification of the file type. The system checks the file's "magic numbers" (the first few bytes of the file) to confirm it matches its claimed MIME type and extension (e.g., ensuring a `.pdf` file is structurally a valid PDF). This prevents attacks based on file type spoofing.
    \item \textbf{Content and Metadata Extraction:} The system then parses the document to extract all textual content. This includes the main body text, comments, annotations, and all metadata fields (e.g., author, title), as any of these can be used to hide prompts.
    \item \textbf{Optical Character Recognition (OCR):} For documents containing images or for non-searchable PDFs (i.e., scans), an OCR engine is applied to convert any text found within images into machine-readable strings. This is critical for detecting instructions hidden in screenshots or diagrams.
    \item \textbf{Macro and Script Analysis:} For file types that support executable code (e.g., Microsoft Office macros), a static analysis engine inspects for suspicious scripts. By default, all active content is stripped or the document is rejected entirely.
    \item \textbf{Unified Text Analysis:} Similar to the audio pipeline, all extracted text from all parts of the document is aggregated into a single text payload. This payload is then sent to the \textbf{SBL} for a full security assessment, ensuring no hidden prompt can bypass the system's defenses.
\end{itemize}

\subsection{Safety Classifiers and Heuristics}
\label{sec:sandbox-classifiers}

\paragraph{Context-Match} This is a crucial logic step where the results of the automated analysis are compared against the user's accompanying text prompt. A significant contradiction, for instance, a prompt requesting a "professional headshot" accompanied by an image flagged for nudity and containing a real face, triggers an immediate rejection and a severe penalty to the user's Trust Score.

\subsection{Tool and Connector Gating}
\label{sec:sandbox-gating}

\paragraph{Threat Vector} A significant threat in agentic systems is the malicious triggering of external tools or connectors through multimodal inputs. For example, an LLM could interpret a QR code in an uploaded image as an instruction to access a malicious URL, or a spoken command could be misinterpreted to trigger a destructive API call (e.g., "delete my last project"). This represents a form of indirect prompt injection where the payload targets the model's tool-use capabilities.

\paragraph{The Gating Mechanism} The Multimodal Sandbox acts as a critical gatekeeper that decouples the interpretation of multimodal data from the execution of tool calls. Before the interpreted semantic content (e.g., "this image contains a QR code for 'malicious.com'") is passed to the LLM's tool-use decision engine, it must first be authorized by the sandbox. The sandbox employs a strict, policy-based control system. For instance, a policy might enforce that "tool calls initiating financial transactions cannot be triggered solely from the content of an image" or that "spoken commands to delete data must be confirmed via a text-based challenge-response." This creates an essential security checkpoint, preventing multimodal inputs from directly executing privileged operations without explicit validation.

\subsection{Forensic Logging (Append-Only / Tamper-Evident)}
\label{sec:sandbox-logging}

All stages of the analysis, including the hashes, classification scores, and the final verdict, are recorded in an append-only, tamper-evident log. This provides an immutable chain of evidence for investigating repeated misuse attempts and can aid in circumventing forensic analysis evasion techniques \cite{TomOr2025}.

\keepXColumns
\begin{tabularx}{\textwidth}{@{} >{\raggedright\arraybackslash}X >{\raggedright\arraybackslash}X >{\raggedright\arraybackslash}X >{\raggedright\arraybackslash}X >{\raggedright\arraybackslash}X @{}}
    \caption{Summary of the Multimodal Input Sandbox Analysis Pipeline.}
    \label{tab:sandbox_pipeline_summary} \\
    \toprule
    \textbf{Check} & \textbf{Technology/Method} & \textbf{Default Threshold} & \textbf{Error Case Handling} & \textbf{Performance Trade-off} \\
    \midrule
    \endfirsthead % Definiert das Ende des Kopfes für die ERSTE Seite
    
    \multicolumn{5}{l}{\tablename~\thetable\ -- \textit{Continued from previous page}} \\
    \toprule
    \textbf{Check} & \textbf{Technology/Method} & \textbf{Default Threshold} & \textbf{Error Case Handling} & \textbf{Performance Trade-off} \\
    \midrule
    \endhead % Definiert den Kopf für alle FOLGESEITEN
    
    \bottomrule
    \endlastfoot % Definiert den Fuß für die LETZTE Seite

    \textbf{Frame Disassembly} & FFmpeg & N/A & Corrupted video file $\rightarrow$ Reject & I/O bound, negligible CPU cost. \\
    \addlinespace
    \textbf{Perceptual Hashing} & ImageHash (e.g., pHash) & Hamming distance $<$ 5 to known CSAM hash & Match found $\rightarrow$ Hard Block, Log, Report & Low CPU, requires database lookups. \\
    \addlinespace
    \textbf{Face Identification} & InsightFace / MediaPipe & Confidence $>$ 0.95 for real face detection & Real face detected in conjunction with nudity $\rightarrow$ Hard Block & Medium CPU, can be a bottleneck for high-res video. \\
    \addlinespace
    \textbf{Nudity Classification} & NudeNet or equivalent & Nudity score $>$ 0.8 & Nudity detected $\rightarrow$ Flag for Context-Match & High CPU (requires GPU for real-time), main latency source. \\
    \addlinespace
    \textbf{Context-Match} & Internal Logic & Mismatch between prompt intent and analysis flags & Mismatch detected $\rightarrow$ Soft Block, increase Trust Score penalty & Low CPU, relies on outputs from previous stages. \\
    \addlinespace
    \textbf{Forensic Logging} & Append-only, tamper-evident (WORM/Merkle) + key-rotation policy & N/A & Log write failure $\rightarrow$ Fail-safe (block request) & I/O bound, requires secure storage. \\
\end{tabularx}

\subsection{Governance, Privacy, and Regional Compliance}
\label{sec:sandbox-governance}

\paragraph{The Governance Challenge} By its nature, the sandbox must inspect user-uploaded data that may be highly sensitive, including images with identifiable faces, voice recordings, or documents containing personal information. This necessitates a robust governance framework to ensure user privacy and legal compliance are maintained at all times.

\paragraph{Core Principles} The system is designed around several core principles:
\begin{itemize}
    \item \textbf{Data Minimization:} Data is retained only for the minimum duration required for security analysis. Raw files are purged immediately after processing. Only anonymized metadata or cryptographic hashes (for comparison against threat intelligence databases) are retained long-term, subject to strict data retention policies.
    \item \textbf{Regional Compliance (e.g., GDPR):} The architecture is designed to be compliant with regional data protection regulations like the GDPR. This is achieved through mechanisms for ensuring a legal basis for processing (i.e., the legitimate interest of securing the platform), obtaining explicit user consent for more invasive analyses, and supporting data residency to ensure user data does not leave a specified jurisdiction.
    \item \textbf{Audited Human-in-the-Loop:} In cases where automated classifiers flag content for human review, strict protocols are enforced to protect user privacy. Access to the data is tightly controlled, fully logged in the immutable audit trail, and restricted to trained security personnel. This ensures accountability and provides a transparent process for handling sensitive edge cases.
\end{itemize}

\subsection{Limitations of the Sandbox}
\label{sec:sandbox-limitations}

While the Multimodal Sandbox provides a critical layer of defense, it is essential to acknowledge its inherent limitations.
\begin{itemize}
    \item \textbf{Classifier Accuracy:} The system relies on machine learning classifiers (e.g., for nudity or risk detection) that are not infallible. \textbf{False negatives} may occur where a sophisticated adversary crafts an input that evades detection. Conversely, \textbf{false positives} can block legitimate user content, negatively impacting the user experience and contributing to a higher False-Block Rate (FBR). This reflects the ongoing "cat-and-mouse game" of adversarial machine learning.
    \item \textbf{Performance Overhead:} As noted in the conceptual evaluation, the analysis of multimodal data is computationally expensive. Video processing, in particular, introduces significant latency, which may be prohibitive for real-time applications. This presents a fundamental trade-off between the depth of security analysis and the desired application performance.
    \item \textbf{Sophisticated Adversarial Techniques:} The sandbox is designed to counter known attack patterns. However, it may be vulnerable to novel, zero-day adversarial techniques. This could include complex cross-modal attacks, where malicious intent is only revealed by combining information from multiple modalities simultaneously, or metaphorical attacks that abuse abstract concepts to bypass concrete classifiers.
    \item \textbf{Scope of Protection:} The sandbox's protection is focused on the content of multimodal inputs. It does not inherently defend against other attack vectors, such as attacks on the underlying cloud infrastructure, denial-of-service attacks at the network layer, or sophisticated data poisoning attacks targeting the training data of the classifiers themselves.
\end{itemize}
% sections/09_methods.tex

\section{Methods and Algorithms}
\label{sec:methods}

This section provides a more formal description of the core algorithms and heuristics that underpin the Countermind architecture. While a full implementation is beyond the scope of this conceptual paper, these descriptions and pseudocode aim to clarify the operational logic and demonstrate the feasibility of the proposed mechanisms.

\subsection{Formal Description of Core Heuristics and Checks}
\label{sec:methods_formal}

The decision-making processes within Countermind can be described with greater formality to illustrate their deterministic and verifiable nature.
\begin{itemize}
    \item \textbf{Text Crypter Validation:} The validation of an incoming payload is a multi-step process that can be formally described. Let $P_{\text{enc}}$ be the encoded payload, $K_{\text{HMAC}}$ be the shared secret HMAC key, and $T_{\text{current}}$ be the current server time. The validation function $V(P_{\text{enc}}, K_{\text{HMAC}}, T_{\text{current}})$ returns a tuple $(P_{\text{plain}}, \text{verdict})$.
    \begin{enumerate}
        \item \textbf{Precondition:} The payload $P_{\text{enc}}$ must conform to the expected format, consisting of an encoded body $B$ and an appended HMAC signature $S$, such that $P_{\text{enc}} = B \mid S$.
        \item \textbf{HMAC Verification:} Compute $S' = \text{HMAC-SHA256}(B, K_{\text{HMAC}})$. If $S' \neq S$, return FAIL; else continue.
        \item \textbf{Decoding and Time Window Check:} The body $B$ is decoded to reveal the plaintext $P_{\text{plain}}$, internal segment proofs $p_1, \dots, p_n$, and an embedded timestamp $T_{\text{gen}}$. The verdict is FAIL if $|T_{\text{current}} - T_{\text{gen}}| > \Delta T_{\max}$, where $\Delta T_{\max}$ is the maximum allowed time window (e.g., 60 seconds).
        \item \textbf{Internal Proof Validation:} The plaintext $P_{\text{plain}}$ is re-segmented into $s_1, \dots, s_n$. For each segment $s_i$, recompute $p'_i$. If any $p'_i \neq p_i$, return FAIL.
        \item \textbf{Output:} If all checks pass, the function returns $(P_{\text{plain}}, \text{PASS})$; otherwise, it returns $(\text{null}, \text{FAIL})$. This process can be related to formal verification techniques that use mathematical proofs to ensure system correctness under specific constraints.
    \end{enumerate}

    \item \textbf{Trust Score Calculation:} The update of the session-based Trust Score $T_s$ for a session $s$ can be modeled as a recursive function: $T_{s, t+1} = \text{clip}(\beta \cdot T_{s,t} + \sum w_k \cdot s_k - \text{penalties}, 0, 1)$, where $\beta$ is a decay factor, $w_k$ are weights for positive signals $s_k$, and penalties are applied for negative signals. The calibration of these parameters is detailed in the Evaluation Plan.
    \begin{itemize}
        \item If $V(R_t)$ is PASS and $\text{IntentDetector}(R_t)$ is LOW\_RISK, a positive signal is added.
        \item If $V(R_t)$ is FAIL or $\text{IntentDetector}(R_t)$ is HIGH\_RISK, a penalty is applied, causing a rapid decay.
    \end{itemize}

    \item \textbf{PSR Policy Enforcement:} The generation process under PSR can be framed as a constrained optimization problem. Let $\theta$ be the LLM parameters and $x$ be the input prompt. The standard generation process aims to find an output sequence $y = y_1, \dots, y_m$ that maximizes the probability $P(y \mid x; \theta)$. Under PSR, the process is modified. Let $A_l(x, y_{<i})$ be the activation vector at layer $l$ before generating token $y_i$. Let $S_{\text{allowed}}$ be the semantic subspace defined by the allowed clusters for the request $x$. The constrained generation process is:
    \begin{align*}
        \max_{y} \quad & P(y \mid x; \theta) \\
        \text{subject to} \quad & A'_{l}(x, y_{<i}) = \Pi_{S_{\text{allowed}}}(A_{l}(x, y_{<i}))
    \end{align*}
    where $\Pi_{S_{\text{allowed}}}$ is a projection operator that ensures the activation vector remains within the allowed subspace, and generation proceeds using the projected activation $A'_l$. This effectively prunes the model's search space of possible next tokens.
\end{itemize}

\subsection{Pseudocode for Critical Decision Pathways}
\label{sec:methods_pseudocode}

To further clarify the system's logic, the following pseudocode illustrates the main request handling flow \cite{Russell2020}.

\begin{lstlisting}[language=Python, caption={High-level pseudocode for the Countermind request handling logic.}, label={lst:request_handling}]
# Define constants and thresholds
SOFT_LOCK_THRESHOLD = 0.4
SEVERE_PENALTY = 0.5

# Define custom exceptions
class PayloadIntegrityError(Exception): pass
class RequestSyntaxError(Exception): pass
class SandboxError(Exception): pass
class RoutingError(Exception): pass

def handle_request(raw_request: dict) -> dict:
    """
    Handles an incoming request, validates it, and routes it.
    Returns: A dictionary representing the response.
    Raises: Various exceptions for different failure modes.
    """
    try:
        # 1. \textbf{SBL}: Decompose and Validate
        origin, meta, payload = decompose_request(raw_request)

        # Layer 1: Syntactic Gate
        if not byte_gate.validate_syntax(payload):
            raise RequestSyntaxError("Invalid request format.")

        # Layer 1: Cryptographic Validation (via Text Crypter)
        plaintext, integrity_ok = text_crypter.decode_and_verify(payload)
        if not integrity_ok:
            trust_scaler.apply_penalty(origin.session_id, SEVERE_PENALTY)
            raise PayloadIntegrityError("Request integrity check failed. Signal decay.")

        # Layers 2 & 3: Routing and Trust Scaling
        route = base_table_router.get_route(meta, plaintext)
        trust_scaler.update_score(origin.session_id, plaintext)

        # Check for Soft-Lock condition
        current_trust_score = trust_scaler.get_score(origin.session_id)
        if current_trust_score < SOFT_LOCK_THRESHOLD:
            log_event(origin, "SOFT_LOCK_ACTIVATED")
            return soft_lock_engine.generate_degradation_response()

        # 2. Route to appropriate handler
        if route == "ROUTE_TO_CORE_LLM":
            # 3. Parameter-Space Restriction
            psr_policy = psr_policy_engine.get_policy(plaintext)
            # 4. Secure Core Logging
            secure_core.log_interaction(origin, plaintext, psr_policy)
            # 5. Constrained Generation
            return core_llm.generate_with_psr(plaintext, psr_policy)
            
        elif route == "ROUTE_TO_MULTIMODAL_SANDBOX":
            sandbox_verdict = multimodal_sandbox.process(payload)
            if sandbox_verdict == "FAIL":
                trust_scaler.apply_penalty(origin.session_id, SEVERE_PENALTY)
                raise SandboxError("Multimodal content violates policy.")
            else:
                # Logic for passing sanitized multimodal data to LLM
                return handle_sanitized_multimodal(payload)

        elif route == "ROUTE_TO_ADMIN_INTERFACE":
            # Handle requests for privileged operations (requires separate auth)
            return handle_admin_request(payload)
            
        else:
            raise RoutingError("Unknown request type.")

    except (RequestSyntaxError, PayloadIntegrityError, SandboxError, RoutingError) as e:
        log_event(origin, type(e).__name__, str(e))
        # Return a generic error response
        return {"status": "error", "message": str(e)}
\end{lstlisting}
% sections/10_evaluation_plan.tex

\section{Evaluation Plan}
\label{sec:evaluation}

As Countermind is a conceptual architecture, this section outlines a proposed protocol for its empirical evaluation. The goal is to demonstrate its effectiveness against various attack vectors and to quantify its performance trade-offs. The evaluation is structured around the core metrics defined in Section \ref{sec:threat_model_metrics}.

\subsection{Proposed Protocol, Baselines, and Measurement Setup}
\label{sec:eval_protocol}

A rigorous evaluation would require a standardized setup to ensure reproducibility and fair comparison. Since these tests have not yet been conducted, this section describes the intended methodology.
\begin{itemize}
    \item \textbf{Models and Datasets:} The protocol would be executed against a diverse set of state-of-the-art LLMs (e.g., Llama 3.1, GPT-4o, Claude 3.5) to assess generalizability. Evaluation datasets would include standard jailbreak suites (e.g., AdvBench), indirect prompt injection corpora, and custom datasets for the attacks described in this paper.
    \item \textbf{Hardware:} All tests would be conducted on a standardized hardware platform (e.g., a cloud instance with NVIDIA H100 GPUs) to ensure consistent performance measurements.
    \item \textbf{Measurement Setup:} The metrics would be measured as follows:
    \begin{itemize}
        \item \textbf{ASR:} An automated judge LLM, supplemented by human review for ambiguous cases, would classify the responses to a benchmark set of adversarial prompts as either successful or failed attacks. Confidence intervals would be calculated for all results.
        \item \textbf{Abstention \& FBR:} These would be measured against curated datasets containing both harmful and benign prompts, respectively.
        \item \textbf{Latency:} The end-to-end response time for each request would be measured, from API call to response receipt, and averaged over multiple runs. This is particularly important for understanding the performance cost of each defensive layer, as components like input filters, cryptographic checks, and output validation all contribute to latency overhead \cite{Shang2025b}.
    \end{itemize}
    \item \textbf{Baselines:} The performance of the full Countermind architecture would be compared against several baselines:
    \begin{enumerate}
        \item \textbf{No Defense:} The raw, unprotected base LLM, to establish a worst-case security baseline.
        \item \textbf{Standard Guardrails:} A configuration designed to mimic typical industry guardrail solutions. This would involve implementing input keyword filtering, output moderation using a classifier, and basic topic control. This provides a benchmark against the current state of the art in reactive defenses.
        \item \textbf{Countermind (Ablated):} Various configurations of Countermind with key components disabled (e.g., Byte-Gate only, +Text-Crypter, +PSR). This allows for an ablation study to measure the specific contribution of each architectural layer to the overall security and performance profile \cite{Meyes2019}.
    \end{enumerate}
\end{itemize}

\subsection{Analysis of Attack Scenarios}
\label{sec:eval_scenarios}

The evaluation would test the system against a wide range of attacks, including those from standard benchmarks and creative, semantically complex attacks.
\begin{itemize}
    \item \textbf{Poetic Non-Instructions Attack:} An attack where malicious instructions are camouflaged within poetic or metaphorical language.
    \begin{itemize}
        \item \textbf{Countermind's Defense:} The primary defense is two-fold. First, the Intent-Detector in the \textbf{SBL} would flag a significant mismatch between the declared metadata and the poetic, instructional nature of the payload. Second, even if the request passes the perimeter, the PSR's strict thematic isolation would prevent the model from acting on the instructions, as the Poetry cluster would not be granted SYNTH rights for system commands.
    \end{itemize}
    \item \textbf{Morphological Attacks:} Attacks that use non-standard character encodings, homoglyphs, or other morphological tricks to obfuscate malicious keywords.
    \begin{itemize}
        \item \textbf{Countermind's Defense:} These attacks would be defeated at the earliest possible stage by the \textbf{Byte-Gate}. Its strict character allowlist, designed to accept only the limited Base62 alphabet of the Text Crypter's output, would immediately reject any input containing such characters.
    \end{itemize}
    \item \textbf{Correction Exploit:} A multi-step social engineering attack where the adversary first submits a query with a deliberate "mistake," then follows up with a "correction" that contains the malicious payload.
    \begin{itemize}
        \item \textbf{Countermind's Defense:} This is a behavioral attack that would be caught by the stateful components of the architecture. The \textbf{Trust-Scaler} would detect the inconsistent and unusual interaction pattern, leading to a penalty on the session's Trust Score. Furthermore, the \textbf{Context-Delta Sentinel (CDS)} would flag the abrupt and significant semantic shift between the initial query and the "correction" as a high-risk anomaly.
    \end{itemize}
\end{itemize}
The following tables present illustrative and conceptual results, representing the \textit{expected outcomes} of the proposed evaluation protocol.

\keepXColumns
\begin{tabularx}{\textwidth}{@{} >{\raggedright\arraybackslash}X c c c c @{}}
    \caption{Conceptual Results: Attack Success Rate (ASR) Comparison (Lower is Better).}
    \label{tab:asr_results} \\
    \toprule
    \textbf{Attack Type} & 
    \textbf{\begin{tabular}{@{}c@{}}Baseline ASR \\ (No Defense)\end{tabular}} & 
    \textbf{\begin{tabular}{@{}c@{}}Baseline ASR \\ (Std. Guardrails)\end{tabular}} & 
    \textbf{\begin{tabular}{@{}c@{}}Countermind \\ ASR\end{tabular}} & 
    \textbf{\begin{tabular}{@{}c@{}}$\Delta$ vs. \\ Guardrails\end{tabular}} \\
    \midrule
    \endfirsthead
    
    \multicolumn{5}{l}{\tablename~\thetable\ -- \textit{Continued from previous page}} \\
    \toprule
    \textbf{Attack Type} & 
    \textbf{\begin{tabular}{@{}c@{}}Baseline ASR \\ (No Defense)\end{tabular}} & 
    \textbf{\begin{tabular}{@{}c@{}}Baseline ASR \\ (Std. Guardrails)\end{tabular}} & 
    \textbf{\begin{tabular}{@{}c@{}}Countermind \\ ASR\end{tabular}} & 
    \textbf{\begin{tabular}{@{}c@{}}$\Delta$ vs. \\ Guardrails\end{tabular}} \\
    \midrule
    \endhead
    
    \bottomrule
    \endlastfoot

    Direct Prompt Injection & 95\% & 40\% & $<$ 1\% & -39\% \\
    Jailbreak (Role-Playing) & 80\% & 30\% & 5\% & -25\% \\
    Indirect Prompt Injection & 70\% & 60\% & 10\% & -50\% \\
    Multimodal Attack (Typographic) & 85\% & 80\% & $<$ 5\% & -75\% \\
    Context Poisoning (Long-term) & 60\% & 55\% & $<$ 5\% & -50\% \\
\end{tabularx}

\subsection{Performance Overhead Analysis}
\label{sec:eval_performance}

A critical aspect of the evaluation is to quantify the performance cost of the added security layers.

\keepXColumns
\begin{tabularx}{\textwidth}{@{} >{\raggedright\arraybackslash}X c c c @{}}
    \caption{Conceptual Results: Performance and Latency Overhead.}
    \label{tab:latency_results} \\
    \toprule
    \textbf{Configuration} & \textbf{Avg. Latency \newline per Turn (ms)} & \textbf{Latency \newline Overhead} & \textbf{Throughput \newline (req/sec)} \\
    \midrule
    \endfirsthead

    \multicolumn{4}{l}{\tablename~\thetable\ -- \textit{Continued from previous page}} \\
    \toprule
    \textbf{Configuration} & \textbf{Avg. Latency \newline per Turn (ms)} & \textbf{Latency \newline Overhead} & \textbf{Throughput \newline (req/sec)} \\
    \midrule
    \endhead

    \bottomrule
    \endlastfoot

    Baseline (No Defense) & 300 & 0\% & 100 \\
    + \textbf{SBL} & 350 & +16.7\% & 85 \\
    + Multimodal Sandbox (Image) & 650 & +116.7\% & 46 \\
    + PSR & 400 & +33.3\% & 75 \\
    Countermind (Full System, Text) & 450 & +50\% & 66 \\
\end{tabularx}

As the table illustrates, the most significant latency overhead is introduced by the Multimodal Sandbox, due to the computationally intensive nature of tasks like nudity classification on high-resolution images. The core text-based defenses \textbf{SBL and PSR} introduce a more moderate but still significant overhead. This highlights the trade-off between security and performance that must be considered during deployment.

\subsection{Conceptual Ablation Study: Contribution of Individual Layers}
\label{sec:eval_ablation}

An ablation study is a standard methodology for understanding the contribution of individual components to a complex system \cite{Meyes2019}. A conceptual ablation of Countermind reveals the critical role of each layer in its defense-in-depth strategy.
\begin{itemize}
    \item \textbf{Without Semantic Boundary Logic (SBL)} The system would be immediately vulnerable to all forms of syntactic and structural attacks. The absence of the Text Crypter would reopen the entire plaintext prompt injection attack surface, rendering the deeper layers largely ineffective against these common threats. The ASR for direct prompt injections would likely revert to the levels of standard guardrails or worse.
    \item \textbf{Without Parameter-Space Restriction (PSR):} The system could block many attacks at the perimeter but would remain vulnerable to sophisticated semantic attacks, zero-day jailbreaks that use novel language, and dangerous emergent behaviors. An adversary could craft a syntactically valid and cryptographically signed prompt that still co-opts the model's internal reasoning to produce a harmful synthesis of information.
    \item \textbf{Without the Secure, Self-Regulating Core:} The system would be static. It could defend against known attacks but would be unable to learn from new patterns observed in the wild. Over time, its defenses would become obsolete as adversaries develop new techniques to bypass its fixed rules. The lack of an immutable audit log would also make forensic analysis and incident response significantly more difficult.
\end{itemize}
This analysis demonstrates that the strength of the Countermind architecture lies not in any single component, but in their synergistic integration. Each layer addresses a different class of threats, and their combination provides a resilient, layered defense.

\begin{figure}[H]
    \centering
    \includegraphics[width=0.9\textwidth]{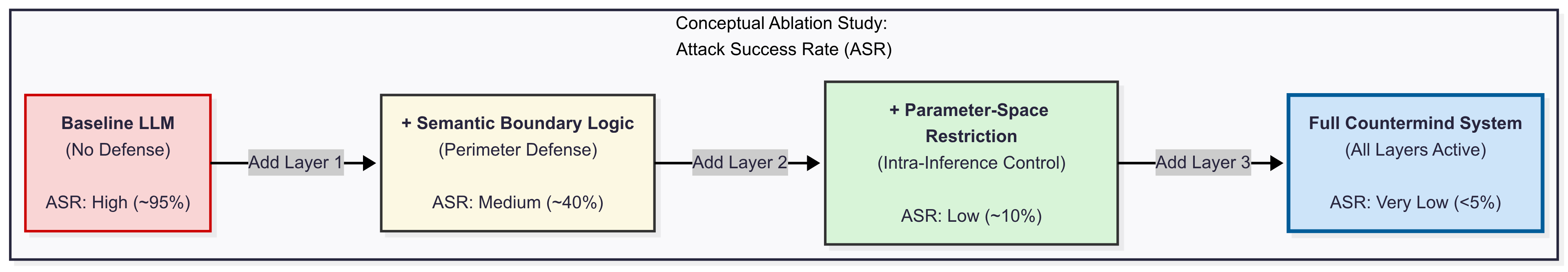}
    \caption{Conceptual Ablation Study - ASR vs. Enabled Components.}
    \label{fig:ablation_study}
\end{figure}
% sections/11_discussion.tex

\section{Discussion}
\label{sec:discussion}

The Countermind architecture represents a departure from current LLM security practices. This section discusses the broader impact of this approach, its applicability to emerging AI paradigms, key operational considerations, and how it compares to existing defensive strategies.

\subsection{Impact and Generalizability to Agents, Tools, and RAG}
\label{sec:discussion_impact}

The principles of Countermind are not limited to simple chatbot applications but are particularly well-suited for securing the next generation of more complex, agentic systems.
\begin{itemize}
    \item \textbf{Agents and Tool Use:} The Parameter-Space Restriction (PSR) framework provides a natural and robust mechanism for governing an agent's use of external tools. Each available tool or API can be mapped to a dedicated semantic cluster (e.g., \texttt{Tools.Email.Send}, \texttt{Tools.Database.Query}). The agent can only invoke a tool if the current request grants it the necessary rights (e.g., SYNTH) for that specific cluster. This prevents common agent vulnerabilities where a hijacked agent is coerced into using a tool for an unauthorized purpose. For example, a prompt injection attack might try to make a customer service agent call the \texttt{delete\_user\_account} tool, but if the PSR policy for that interaction does not include the \texttt{Tools.Admin.Delete} cluster, the action would be blocked at the activation level before the tool is ever called.
    \item \textbf{Retrieval-Augmented Generation (RAG):} RAG systems, which augment LLM prompts with information retrieved from external knowledge bases, are a primary vector for indirect prompt injection. An attacker can poison the knowledge base with documents containing hidden malicious instructions. Countermind addresses this in two ways. First, the OMP decomposition allows the system to treat the user's query and the retrieved documents as separate entities with different trust levels. Second, Semantic Zoning can place all retrieved content into a specific context zone (e.g., \texttt{RAG.RetrievedData}) with strict, READ-ONLY rights. This policy is enforced at the PSR level, ensuring that \texttt{RAG.RetrievedData} receives permanent READ-ONLY, with SYNTH/EVAL=deny. This ensures that the LLM can use the information from the documents to formulate its answer but cannot interpret any text within those documents as an executable instruction.
\end{itemize}

\subsection{Operational Considerations: Tuning, Logging, and Fail-Safe vs. Fail-Open Design}
\label{sec:discussion_operational}

Deploying a system like Countermind in a production environment requires careful consideration of several operational aspects.
\begin{itemize}
    \item \textbf{Tuning and Usability:} The effectiveness of the architecture depends on the careful tuning of its various thresholds, such as the Trust Score for activating the Soft-Lock, and the confidence scores for the Multimodal Sandbox classifiers. Overly aggressive settings will lead to a high False-Block Rate (FBR), frustrating legitimate users and reducing the application's utility. Conversely, settings that are too lenient will fail to stop sophisticated attacks. Achieving the right balance requires an iterative process of testing, monitoring, and adjustment, guided by the core metrics of ASR and FBR.
    \item \textbf{Logging and Forensics:} The immutable audit log is not just a security feature but a critical operational tool. It provides the necessary data for debugging system failures, conducting forensic investigations after a security incident, demonstrating compliance with regulatory requirements, and providing the training data for the Learning Security Core. The integrity and availability of this log are paramount to the system's long-term health and adaptability.
    \item \textbf{Fail-Safe vs. Fail-Open Design:} A critical architectural decision for any security system is its behavior in the event of an internal failure, such as a component crash or a lost connection to a database. The two primary design philosophies are fail-safe (also known as fail-closed) and fail-open \cite{Anderson2008}.
    \begin{itemize}
        \item \textbf{Fail-Safe (Fail-Closed):} In this mode, the system defaults to its most secure state, which typically means blocking the request. This prioritizes security over availability. For example, if the PSR policy engine fails, a fail-safe design would reject all requests rather than allowing them to proceed to the LLM without governance.
        \item \textbf{Fail-Open:} In this mode, the system defaults to an operational state, bypassing the failed security check to maintain availability. This might be appropriate for non-critical systems where uptime is more important than perfect security.
        \item \textbf{Degraded-Safe Mode:} A third path exists where, upon failure of a non-critical component (e.g., the learning core), the system enters a degraded-safe mode. In this state, it might only permit low-risk operations (e.g., READ rights) while disabling more privileged ones (SYNTH, CROSS) until the component is restored. Countermind's default posture is designed to be fail-safe. Given the potential for LLMs to cause significant harm, the architecture prioritizes preventing a security failure over ensuring 100\% availability.
    \end{itemize}
\end{itemize}

\subsection{Comparison to Existing Paradigms: Guardrails, RLHF, and Constitutional AI}
\label{sec:discussion_comparison}

Countermind builds upon concepts from existing security paradigms but differs in its fundamental approach.
\begin{itemize}
    \item \textbf{Comparison to Guardrails:} Guardrail systems provide a powerful framework for defining and enforcing programmable safety rules for LLMs. They typically operate at the application layer, evaluating prompts and responses against a defined set of policies. Countermind complements this approach but operates at a deeper, more architectural level. The \textbf{SBL}, with its cryptographic Text Crypter, provides a structural defense that is not based on programmable rules but on verifiable mathematical properties of the input. The PSR operates at the sub-symbolic, activation level, providing a form of control that is closer to the model's core logic than a typical guardrail system. The measurable difference lies in the point of intervention: pre-inference structural validation versus post hoc content filtering, which is expected to yield a lower ASR for structural attacks and a different latency profile, as outlined in our evaluation plan.
    \item \textbf{Comparison to RLHF and Constitutional AI:} Reinforcement Learning from Human Feedback (RLHF) and its derivative, Constitutional AI (CAI), are the dominant methods for aligning LLM \textit{behavior} \cite{Bai2022}. These techniques work by training a preference model on human or AI-generated feedback and then using that model as a reward function to fine-tune the LLM to produce more desirable outputs. This is a powerful but fundamentally \textbf{reactive} training process; it teaches the model to prefer safe outputs over unsafe ones based on past examples. Countermind, in contrast, is a \textbf{proactive} inference-time control mechanism. PSR does not retrain the model but constrains its generative process in real-time. While CAI provides the LLM with a set of principles to \textit{follow}, Countermind's Secure Core and PSR provide an architectural framework to \textit{enforce} those principles, regardless of the model's learned behavior. This makes it potentially more robust against novel attacks that the model was not trained to handle.
    \item \textbf{Comparison to Point-Defenses:} The current research landscape is populated with many "point-defenses," which are specific algorithms designed to detect a particular type of attack, such as a classifier for prompt injection. While valuable, these solutions often lack generalizability. Countermind is proposed as a holistic, defense-in-depth architecture. It does not rely on a single mechanism but integrates multiple, diverse defenses (cryptographic, syntactic, semantic, behavioral, and sub-symbolic) to provide layered and resilient protection against a wide spectrum of threats.
\end{itemize}
% sections/12_limitations.tex

\section{Limitations and Future Work}
\label{sec:limitations}

While the Countermind architecture presents a comprehensive conceptual framework for LLM security, it is important to acknowledge its limitations and areas for future research.
\begin{itemize}
    \item \textbf{Implementation Complexity:} The most significant limitation is the engineering complexity of implementing the proposed system. The \textbf{SBL} and Multimodal Sandbox are non-trivial but build upon existing technologies. The Parameter-Space Restriction (PSR) mechanism, however, requires deep integration with the LLM's inference engine to manipulate activations during the forward pass. This is a substantial technical challenge that would require close collaboration with model developers or advanced expertise in modifying open-source model architectures.
    \item \textbf{Key Management:} The security of the Text Crypter relies on the secure provisioning, rotation, and revocation of HMAC keys for legitimate clients. A robust key management infrastructure is a critical and complex prerequisite for this architecture.
    \item \textbf{Semantic Cluster Definition:} The effectiveness of PSR is highly dependent on the quality and granularity of the defined semantic clusters. The process of identifying and delineating these clusters within a model's activation space is a complex research problem in itself. It would likely require a combination of unsupervised clustering techniques, interpretability tools, and significant domain expertise to create a meaningful and robust cluster map. Future work should explore semi-automated methods for defining and maintaining these semantic taxonomies.
    \item \textbf{Performance Overhead and User Friction:} As shown in the conceptual evaluation, the security layers introduce a non-negligible latency overhead. This may be prohibitive for applications requiring near-instantaneous responses. Furthermore, strict validation mechanisms, such as a short time-window for the Text Crypter, could lead to a high FBR due to network latency or clock skew, creating significant user friction.
    \item \textbf{Multimodal False Positives/Negatives:} The Multimodal Sandbox relies on classifiers that are not perfect. False negatives could allow harmful content to pass, while false positives could block legitimate user content, impacting usability. The plan for measuring these error rates is crucial for understanding the practical viability of this component.
    \item \textbf{Security of the Architecture Itself:} The framework does not inherently defend against attacks on the underlying infrastructure, such as the host operating system or hypervisor. A compromise of the trusted computing base would undermine the entire architecture.
    \item \textbf{Evasion of High-Level Controls:} While PSR is designed to be robust against many semantic attacks, it is conceivable that a highly sophisticated adversary could develop novel techniques to bypass the cluster-based controls. For example, an attacker might find ways to express a forbidden concept using only the representations available in a set of allowed clusters. The resilience of the PSR governance model against such advanced, second-order semantic attacks is an open area for investigation.
\end{itemize}
Future work should focus on building a proof-of-concept implementation of the Countermind architecture to empirically validate the conceptual claims made in this paper. This would involve tackling the challenges of PSR implementation and cluster definition, and conducting rigorous, large-scale evaluations against a comprehensive suite of adversarial benchmarks.
% sections/13_ethics.tex

\section{Ethics Statement \& Dual-Use Considerations}
\label{sec:ethics}

The research and concepts presented in this paper are intended to advance the field of defensive AI security and promote the development of safer, more trustworthy LLM systems. The following ethical considerations have been taken into account.
\begin{itemize}
    \item \textbf{Dual-Use:} All security research has a potential dual-use nature; techniques developed for defense can sometimes inform the development of new attacks. The focus of this work is exclusively on hardening AI systems. The architectural principles are described at a level intended to guide secure system design without providing a step-by-step blueprint for a specific exploitable vulnerability in any existing system.
    \item \textbf{Responsible Disclosure:} This paper does not contain details of any specific, zero-day vulnerabilities discovered in any commercial or open-source LLM products. The attack scenarios discussed are either conceptual or based on publicly known attack classes. The goal is to foster a culture of proactive security design, in line with the principles of responsible disclosure.
    \item \textbf{Benign Payloads:} All examples of adversarial prompts or malicious payloads described in this paper are for illustrative and evaluation purposes only. They are designed to be conceptually representative of real threats without containing actually harmful, dangerous, or illegal content.
    \item \textbf{User Notice and Governance for Soft-Lock:} The "Soft-Lock" or graceful degradation response mechanism presents an ethical challenge as it involves a level of deception towards the user. A User Notice Policy must be established to govern its use. This policy should define:
    \begin{itemize}
        \item \textbf{Transparency:} The system's terms of use should clearly state that interactions may be monitored for security purposes and that the service may be gracefully degraded if malicious intent is detected.
        \item \textbf{Audit and Appeal:} All Soft-Lock activations must be recorded in the immutable audit log. A clear process must be available for users to appeal a restriction if they believe it was applied in error (a false positive), triggering a human-in-the-loop review.
        \item \textbf{Proportionality:} The duration of a Soft-Lock should be proportionate to the detected risk, with policies for automatic expiration and review to mitigate the impact of false positives.
    \end{itemize}
    \item \textbf{Compliance for Multimodal Analysis:} The analysis of user-uploaded content in the Multimodal Sandbox, particularly involving pHash databases or facial recognition, must be conducted in strict compliance with legal and ethical standards. This includes a clear legal basis for processing, adherence to jurisdictional regulations (e.g., GDPR), minimal data retention policies for sensitive content, and a human review process for handling ambiguous cases or false positives. The system should perform classification and recognition, not biometric identification, unless a documented legal basis is provided.
\end{itemize}
% sections/14_declarations.tex

\section*{Declarations and Statements}
\addcontentsline{toc}{section}{Declarations and Statements}
\label{sec:declarations}

\paragraph{Use of AI Tools:}
Large Language Models were utilized as research assistants during the preparation of this manuscript. Their use included summarizing academic papers, assisting with the translation of conceptual notes, and providing grammar and style corrections. All core concepts, architectural designs, analyses, and the final written text were directed, authored, and verified by the human author. No section of this paper consists of unverified, directly generated output from an AI model.

\paragraph{Conflicts of Interest:}
The author declares no conflicts of interest.

\paragraph{Funding:}
This research received no specific grant from any funding agency in the public, commercial, or not-for-profit sectors.

\paragraph{Data and Code Availability:}
This paper describes a conceptual architecture. No new datasets were generated.
% sections/15_acknowledgments.tex

\section*{Acknowledgments}
\addcontentsline{toc}{section}{Acknowledgments}
\label{sec:acknowledgments}

The author wishes to thank the open-source community and the many researchers whose work in LLM security, interpretability, and alignment provided the foundation for the concepts explored in this paper. This work is distributed under the Creative Commons Attribution 4.0 International (CC BY 4.0) license.

%========= LITERATURVERZEICHNIS =========
\bibliographystyle{unsrt} 
\bibliography{references} 

\begin{thebibliography}{10}

\bibitem{AguileraMartinez2025}
Fabricio Aguilera-Mart{\'\i}nez and Fernando Berzal.
\newblock {LLM} security: Vulnerabilities, attacks, defenses, and
  countermeasures, 2025.

\bibitem{Zou2023a}
Andy Zou, Long Phan, Sarah Chen, Keyan Nasseri, Oliver Zhang, Angela Jiang,
  Mantas Mazeika, Ann-Kathrin Dombrowski, Dan Hendrycks, Jacob Steinhardt, and
  Avital Oliver.
\newblock Representation engineering: A top-down approach to {AI} transparency,
  2023.

\bibitem{Rossi2024}
Simone Rossi, Andrea~Mauro Michel, Raghava~Rao Mukkamala, and Jason~Bennett
  Thatcher.
\newblock {An Early Categorization of Prompt Injection Attacks on Large
  Language Models}, 2024.

\bibitem{Xu2025}
Wenrui Xu and Keshab~K. Parhi.
\newblock A survey of attacks on large language models, 2025.

\bibitem{Liu2024a}
Xiaogeng Liu, Zhiyuan Yu, Yizhe Zhang, Ning Zhang, and Chaowei Xiao.
\newblock {Automatic and Universal Prompt Injection Attacks against Large
  Language Models}, 2024.

\bibitem{Shang2025a}
Zhengchun Shang and Wenlan Wei.
\newblock Evolving security in {LLM}s: A study of jailbreak attacks and
  defenses, 2025.

\bibitem{Chiu2025}
Chun~Wai Chiu, Linghan Huang, Bo~Li, Huaming Chen, and Kim{-}Kwang~Raymond
  Choo.
\newblock {``Do as I Say, Not as I Do''}: A semi-automated approach for
  jailbreak prompt attacks against multimodal {LLM}s, 2025.

\bibitem{Cheng2024}
Hao Cheng, Erjia Xiao, Jindong Gu, Le~Yang, Jinhao Duan, Jize Zhang, Jiahang
  Cao, Kaidi Xu, and Renjing Xu.
\newblock Unveiling typographic deceptions: Insights into the typographic
  vulnerability in large vision{-}language models, 2024.

\bibitem{Kong2025}
Dezhang Kong, Shi Lin, Zhenhua Xu, Zhebo Wang, Minghao Li, Yufeng Li, Yilun
  Zhang, Hujin Peng, Zeyang Sha, Yuyuan Li, Changting Lin, Xun Wang, Xuan Liu,
  Ningyu Zhang, Chaochao Chen, Muhammad~Khurram Khan, and Meng Han.
\newblock {A Survey of {LLM}-Driven {AI} Agent Communication: Protocols,
  Security Risks, and Defense Countermeasures}, 2025.

\bibitem{LiFung2025}
Miles~Q. Li and Benjamin C.~M. Fung.
\newblock Security concerns for large language models: A survey, 2025.

\bibitem{Zhang2025}
Chi Zhang, Changjia Zhu, Junjie Xiong, Xiaoran Xu, Lingyao Li, Yao Liu, and
  Zhuo Lu.
\newblock Guardians and offenders: A survey on harmful content generation and
  safety mitigation of {LLM}, 2025.

\bibitem{Chen2025SafetyConstraints}
Xin Chen, Yarden As, and Andreas Krause.
\newblock Learning safety constraints for large language models, 2025.

\bibitem{Yi2024}
Sibo Yi, Yule Liu, Zhen Sun, Tianshuo Cong, Xinlei He, Jiaxing Song, Ke~Xu, and
  Qi~Li.
\newblock Jailbreak attacks and defenses against large language models: A
  survey, 2024.

\bibitem{Schulhoff2025}
Nishant Balepur, Rachel Rudinger, and Jordan~Lee Boyd-Graber.
\newblock Which of these best describes multiple choice evaluation with {LLM}s?
  a) forced b) flawed c) fixable d) all of the above, 2025.

\bibitem{Wehner2025}
Jan Wehner, Sahar Abdelnabi, Daniel Tan, David Krueger, and Mario Fritz.
\newblock Taxonomy, opportunities, and challenges of representation engineering
  for large language models, 2025.

\bibitem{Turner2023}
Alexander~Matt Turner, Lisa Thiergart, Gavin Leech, David Udell, Juan~J.
  Vazquez, Ulisse Mini, and Monte MacDiarmid.
\newblock Steering language models with activation engineering, 2023.

\bibitem{Lee2024PRCAS}
Bruce~W. Lee, Inkit Padhi, Karthikeyan~Natesan Ramamurthy, Erik Miehling,
  Pierre Dognin, Manish Nagireddy, and Amit Dhurandhar.
\newblock Programming refusal with conditional activation steering, 2024.

\bibitem{Bai2022}
Yuntao Bai, Saurav Kadavath, Sandipan Kundu, Amanda Askell, Jackson Kernion,
  Tom Conerly, Emily Chen, Andy Jones, Deep Ganguli, Anna Goldie, Ben Mann,
  Nelson Joseph, Sam McCandlish, Kamaldeep Singh, Jared Kaplan, and Dario
  Amodei.
\newblock Constitutional ai: Harmlessness from {AI} feedback, 2022.

\bibitem{Chen2025}
Tiejin Chen, Pingzhi Li, Kaixiong Zhou, Tianlong Chen, and Hua Wei.
\newblock Unveiling privacy risks in multi{-}modal large language models:
  Task{-}specific vulnerabilities and mitigation challenges.
\newblock In {\em Findings of the Association for Computational Linguistics:
  ACL 2025}, 2025.

\bibitem{Wu2024}
Tingmin Wu, Shuiqiao Yang, Shigang Liu, David Nguyen, Seung Jang, and Alsharif
  Abuadbba.
\newblock Threatmodeling{-}{LLM}: Automating threat modeling using large
  language models for banking system, 2024.

\bibitem{Yang2024a}
Yuchen Yang, Yiming Li, Hongwei Yao, Bingrun Yang, Yiling He, Tianwei Zhang,
  Dacheng Tao, and Zhan Qin.
\newblock Shadowcode: Towards (automatic) external prompt injection attack
  against code {LLM}s, 2024.

\bibitem{Zou2023b}
Andy Zou, Zifan Wang, Nicholas Carlini, Milad Nasr, J.~Zico Kolter, and Matt
  Fredrikson.
\newblock Universal and transferable adversarial attacks on aligned language
  models, 2023.

\bibitem{Wu2025b}
Liangxuan Wu, Chao Wang, Tianming Liu, Yanjie Zhao, and Haoyu Wang.
\newblock From assistants to adversaries: Exploring the security risks of
  mobile {LLM} agents, 2025.

\bibitem{Freenor2025}
Michael Freenor, Lauren Alvarez, Milton Leal, Lily Smith, Joel Garrett,
  Yelyzaveta Husieva, Madeline Woodruff, Ryan Miller, Erich Kummerfeld, Rafael
  Medeiros, and Sander Schulhoff.
\newblock Prompt optimization and evaluation for {LLM} automated red teaming,
  2025.

\bibitem{Wen2024}
Bingbing Wen, Jihan Yao, Shangbin Feng, Chenjun Xu, Yulia Tsvetkov, Bill Howe,
  and Lucy~Lu Wang.
\newblock Know your limits: A survey of abstention in large language models,
  2024.

\bibitem{Li2025b}
Boshi Huang and Fabio~Nonato {de Paula}.
\newblock Defend llms through self-consciousness, 2025.

\bibitem{Shang2025b}
Zhifan Luo, Shuo Shao, Su~Zhang, Lijing Zhou, Yuke Hu, Chenxu Zhao, Zhihao Liu,
  and Zhan Qin.
\newblock Shadow in the cache: Unveiling and mitigating privacy risks of
  {KV}-cache in llm inference, 2025.

\bibitem{LaMalfa2024}
Francesco Periti and Stefano Montanelli.
\newblock {\em {Lexical Semantic Change through Large Language Models: a
  Survey}}.
\newblock PhD thesis, University of Milan, 2024.

\bibitem{Balarabe2025}
T.~Balarabe.
\newblock Understanding {LLM} context windows, tokens, attention, and
  challenges.
\newblock Medium blog post, 2025.
\newblock non-peer-reviewed.

\bibitem{Yang2025}
Yuchen Yang, Qichang Liu, Christopher Brix, Huan Zhang, and Yinzhi Cao.
\newblock Certphash: Towards certified perceptual hashing via robust training.
\newblock In {\em Proceedings of the 34th USENIX Security Symposium}, 2025.

\bibitem{TomOr2025}
Tom Or and Omri Azencot.
\newblock Unraveling hidden representations: A multi{-}modal layer analysis for
  better synthetic content forensics, 2025.

\bibitem{Russell2020}
Stuart Russell and Peter Norvig.
\newblock {\em Artificial Intelligence: A Modern Approach}.
\newblock Pearson, 4th edition, 2020.

\bibitem{Meyes2019}
Richard Meyes, Melanie Lu, Constantin~Waubert de~Puiseau, and Tobias Meisen.
\newblock Ablation studies in artificial neural networks, 2019.

\bibitem{Anderson2008}
Ross~J. Anderson.
\newblock {\em Security Engineering: A Guide to Building Dependable Distributed
  Systems}.
\newblock Wiley, 2nd edition, 2008.

\end{thebibliography}

%========= Appendix =========
\clearpage
\begin{appendices}
\begingroup
\titleformat{\section}[block]{\large\bfseries}{Appendix~\thesection:}{0.6em}{}
\section{Cluster Policy Artifact}\label{app:policy}

{\small
\begin{verbatim}
version: 1
default: deny
clusters:
  READ:
    allow: [literal_lookup]              # retrieval without synthesis
    deny: [code_generation, cross_source_aggregation]
    thresholds: { trust_score_min: 0.70, human_review: false }

  SYNTH:
    allow: [summarize, paraphrase]
    deny: [code_execution, system_calls]
    thresholds: { trust_score_min: 0.80, human_review: true }

  EVAL:
    allow: [static_analysis, constraint_check]
    deny: [external_write]
    thresholds: { trust_score_min: 0.75, human_review: true }

  CROSS:
    allow: [cross_source_compare]
    deny: [freeform_generation, external_post]
    thresholds: { trust_score_min: 0.85, human_review: true }

audit:
  mode: append_only
  fields: [timestamp, user, cluster, decision, reason]
\end{verbatim}
}
\endgroup
\end{appendices}

\end{document}